%% file: SPC_Paper_Monograph.tex
\documentclass[ALICE,manyauthors]{cernphprep}
\usepackage[comma,square,numbers,sort&compress]{natbib}
\usepackage[]{hyperref}
\usepackage{cleveref}
\Crefname{figure}{Fig.}{Figs.}

\usepackage{lineno}
\usepackage{xspace}
\usepackage{enumerate}
\usepackage[dvipsnames]{xcolor}
\usepackage[T1]{fontenc}
\usepackage{orcidlink}

\begin{document}
\input{commands.tex}

%%%%%%%%%%%%%%%  Title page %%%%%%%%%%%%%%%%%%%%%%%%
\begin{titlepage}
% the dates below correspond to CERN approval
% please don't touch: EB chairs will take care
\PHyear{2024}       % required, will be obtained from CERN
\PHnumber{223}      % required, will be obtained from CERN
\PHdate{29 August}  % required, will be obtained from CERN
%%%%%%%%%%%%%%%%%%%%%%%%%%%%%%%%%%%%%%%%%%%%%%%%%%%%

%%% Put your own title + short title here:
%\title{Higher precision measurements on symmetry plane correlations in Pb--Pb collisions at \fivenn}
\title{Higher-order symmetry plane correlations in Pb--Pb collisions at $\sqrt{\mathbf{\textit{s}}_{\mathrm{\textbf{NN}}}}~\mathbf{=}~\mathbf{5.02}$~Te\kern-.1emV\xspace}
\ShortTitle{Higher-order symmetry plane correlations in Pb--Pb collisions}   % appears on left page headers

%%% Do not change the next lines
\Collaboration{ALICE Collaboration\thanks{See Appendix~\ref{app:collab} for the list of collaboration members}}
\ShortAuthor{ALICE Collaboration} % appears on right page headers, do not change

\begin{abstract}

The correlations between event-by-event fluctuations of symmetry planes are measured in Pb--Pb collisions at a centre-of-mass energy per nucleon pair \fivenn~recorded by the ALICE detector at the Large Hadron Collider.
This analysis is conducted using the Gaussian Estimator technique, which is insensitive to biases from correlations between different flow amplitudes.
The study presents, for the first time, the centrality dependence of correlations involving up to five different symmetry planes.
The correlation strength varies depending on the harmonic order of the symmetry plane and the collision centrality.
Comparisons with measurements from lower energies indicate no significant differences within uncertainties.
Additionally, the results are compared with hydrodynamic model calculations. Although the model predictions provide a qualitative explanation of the experimental results, they overestimate the data for some observables. This is particularly true for correlators that are sensitive to the non-linear response of the medium to initial-state anisotropies in the collision system.
As these new correlators provide unique information --- independent of flow amplitudes --- their usage in future model developments can further constrain the properties of the strongly-interacting matter created in ultrarelativistic heavy-ion collisions.

\end{abstract}
\end{titlepage}

\setcounter{page}{2} %please do not remove this line

%%%%%%%%%%%%%%%%%%%%%%%%%%%%%%%%
% begin main text
%%%%%%%%%%%%%%%%%%%%%%%%%%%%%%%%
\section{Introduction}
\label{sec:intro}

One of the most extensively studied phenomena in ultrarelativistic heavy-ion collisions is the collective behaviour of the produced particles~\cite{Ollitrault:1992bk, Heinz:2013th}. The nature of the created system and the interactions among particles involved make collectivity sensitive to all stages of the collision~\cite{Gale:2013da, Luzum:2013yya, Braun-Munzinger:2015hba, Bernhard:2016tnd, Shen:2020mgh, Parkkila:2021tqq, Parkkila:2021yha, ALICE:2022wpn}.
Moreover, previous studies, including Bayesian parameter estimations, have shown that the measurements of the correlations between the charged final-state particles are, in particular, a highly sensitive probe of the transport properties of the quark--gluon plasma (QGP)~\cite{Bernhard:2015hxa, ALICE:2016kpq, Bernhard:2016tnd, ALICE:2017kwu, Bernhard:2019bmu, Parkkila:2021yha, ALICE:2022wpn}.
This extreme state of matter, created during the collision, consists of deconfined quarks and gluons, and exhibits quasi-perfect liquid-like behaviour. 
As the QGP cannot be directly observed and the evolving system is highly complex, it is crucial to develop and measure observables related to the collectivity of the final-state particles, which reflects the properties of the underlying medium. More details about the QGP studies can be found in Refs.~\cite{Adcox:2004mh, Adams:2005dq, Braun-Munzinger:2015hba, Bernhard:2016tnd, Shen:2020mgh, ALICE:2022wpn}.

One observable manifestation of the collective behaviour of the detected particles is their anisotropic azimuthal emission in the transverse plane, known as the anisotropic flow. This phenomenon arises from the medium response to the non-isotropic initial-state geometry of the collision and the initial inhomogeneities in the system's entropy density~\cite{Ollitrault:1992bk, Voloshin:1994mz}.
The azimuthal distribution of the final-state particles $f(\varphi)$ can be represented using a Fourier series as~\cite{Voloshin:1994mz}
\begin{equation}
    f(\varphi) = \frac{1}{2\pi} \left[1 + 2\sum_n v_n \cos[n(\varphi - \Psi_n)]\right],
\end{equation}
where $v_n$ and $\Psi_n$ denote the flow amplitudes and symmetry plane angles of the $n$-th harmonic, respectively.
The measurements and theoretical calculations of these quantities and their complex correlations play an important role in understanding the properties of QGP~\cite{ALICE:2016kpq, ALICE:2017kwu, ALICE:2020sup, ALICE:2021klf, ALICE:2021adw, ALICE:2023wdn, ALICE:2023lwx}.

Flow amplitudes and symmetry planes, as well as their event-by-event fluctuations, are not directly accessible experimentally. Therefore, various approaches were developed to estimate them, such as the event plane method~\cite{E877:1996czs, Poskanzer:1998yz} or the multiparticle correlation techniques~\cite{Borghini:2000sa, Borghini:2001vi, Bilandzic:2010jr, Bhalerao:2011yg, Bilandzic:2013kga}.
The latter, which are the focus of this analysis, rely on the relation between the measured particle azimuthal angles $\varphi$ and the flow degrees of freedom $v_n$ and $\Psi_n$~\cite{Bhalerao:2011yg, Bilandzic:2013kga, Bilandzic:2020csw}
\begin{equation}
    v_{n_1}^{a_1}\cdots v_{n_k}^{a_k}\,e^{i(a_1 n_1\Psi_{n_1}+\cdots+a_k n_k\Psi_{n_k})} = \left<e^{i(n_1\varphi_1+\cdots+n_l\varphi_l)}\right>,
    \label{eq:generalResult}
\end{equation}
where $\langle\ldots\rangle$ indicates an average over the azimuthal angles of distinct combinations of $l$ particles in the same event.
Given that the multiparticle azimuthal correlator contains $k$ ($k\leq 
 l$) different harmonics (counting positive and negative harmonics separately), the positive integers $a_i$ are used to ensure each harmonic $n_i$ appears only once on the left-hand side of Eq.~\eqref{eq:generalResult}.
The order of the multiparticle azimuthal correlator $l$, which corresponds to the number of particles involved, is given by $l = \sum_i a_i$. Further details on the coefficients $a_i$ and their interpretation can be found in Ref.~\cite{ALICE:2023wdn}.

In the last few years, significant progress has been made in the study of the flow amplitudes $v_n$. The flow coefficients themselves were measured for the harmonics from $n = 2$~\cite{STAR:2001ksn, STAR:2002hbo, STAR:2003xyj,Adams:2005dq,Adcox:2004mh,ALICE:2011ab,ATLAS:2018ezv} to $n = 9$~\cite{ALICE:2020sup}. The first experimental studies of the symmetric cumulants (SC) by the ALICE Collaboration~\cite{ALICE:2016kpq, ALICE:2017kwu} showed the existence of non-negligible correlations between two different flow amplitudes $v_m^2$ and $v_n^2$.
This research were extended to include more flow amplitudes~\cite{Mordasini:2019hut, Moravcova:2020wnf, ALICE:2021klf, ALICE:2021adw} and different moments of these amplitudes' distributions~\cite{Moravcova:2020wnf, Bilandzic:2021rgb, ALICE:2021adw, ALICE:2023lwx}.
Comparisons between experimental results and theoretical calculations have demonstrated that the measurements of the SC are more sensitive to the transport properties of the QGP than the individual flow amplitudes~\cite{ALICE:2016kpq}.
Such a conclusion was recently confirmed with sensitivity studies conducted in Bayesian analyses~\cite{Parkkila:2021tqq, Parkkila:2021yha}. These studies proved that flow observables involving higher-order harmonics or cumulants impose more constraints on the initial-state and hydrodynamic model parameters.

Similarly to the flow amplitudes, symmetry planes carry information about both the initial state of the collision and the QGP phase. As such, analysing symmetry planes and their correlations provides additional insights into the QGP produced in those collisions. However, even in the case of collisions with impact parameters of same magnitude in absence of event-by-event fluctuations of the position of participating nucleons, the angle of each symmetry plane $\Psi_n$ measured in the laboratory frame varies from one collision to another due to a different orientation of the impact parameter. This orientation is generally characterised as the angle of the reaction plane ($\Psi_\mathrm{RP}$), which is a plane spanned by the impact parameter and the beam axis.
The reaction plane angle $\Psi_\mathrm{RP}$, therefore, changes from one event to another like $\Psi_\mathrm{RP} \rightarrow \Psi_\mathrm{RP} + \alpha$, where $\alpha$ is a random fluctuation.  
Hence, each individual symmetry plane shifts by $\Psi_n \rightarrow \Psi_n + \alpha$ in the laboratory frame from one collision to another. As a result, the averaged value of an individual symmetry plane measured in the laboratory is zero.
Therefore, the simplest non-trivial observables are the correlated fluctuations of different symmetry planes with respect to each other, as the contribution of the reaction plane fluctuations cancels out. These are quantified by observables such as $\langle \cos[4(\Psi_4-\Psi_2)] \rangle$, which is one example involving the symmetry planes $\Psi_4$ and $\Psi_2$.
The analysis of these symmetry plane correlations (SPC) presents unique challenges. For instance, any expression of SPC has to be rotationally invariant, ensuring that the random event-by-event fluctuations of the impact parameter vector do not affect the measurement. This can be achieved by a suitable choice of the harmonics $n_i$ and coefficients $a_i$ in Eq.~\eqref{eq:generalResult} with the constraint $\sum a_i n_i = 0$. More details on the proper choice of these quantities, which lead to non-trivial SPC, can be found in Appendix~B of Ref.~\cite{Bilandzic:2020csw}.

When applying Eq.~\eqref{eq:generalResult} in experimental measurements, it is not possible to find a set of harmonics such that the left-hand side of Eq.~\eqref{eq:generalResult} only involves symmetry planes without the inclusion of flow amplitudes. As a solution to this, estimators of SPC were developed to remove the contribution of the flow amplitudes when using multiparticle correlation techniques.
One of the most widely used methods to estimate the SPC is the so-called Scalar Product (SP) method~\cite{ATLAS:2014ndd, ALICE:2017fcd}, which was used in the analyses of SPC in heavy-ion collisions at the Large Hadron Collider (LHC) energies by ATLAS~\cite{ATLAS:2014ndd} and ALICE~\cite{ALICE:2017fcd, ALICE:2020sup}. The SP method is defined as
\begin{equation}
    \langle \cos (a_1 n_1 \Psi_{n_1} +\dots+ a_k n_k \Psi_{n_k})\rangle_{\text{SP}} = \frac{\langle v_{n_1}^{a_1}\dots v_{n_k}^{a_k} \cos(a_1 n_1 \Psi_{n_1} +\dots+ a_k n_k \Psi_{n_k})\rangle}{\sqrt{ \langle v_{n_1}^{2a_1} \rangle\dots \langle v_{n_k}^{2a_k}\rangle}}\,,
    \label{eq. SP-Method}
\end{equation}
where both numerator and denominator can be estimated with multiparticle correlation techniques using Eq.~\eqref{eq:generalResult}. In this approach, one divides the numerator, which contains the mean of the product of flow amplitudes, by the product of their respective means in the denominator. Such a ratio, however, introduces a bias in the estimation of SPC as correlations among flow amplitudes imply that $\langle v_{n_1}^{a_1} \dots v_{n_k}^{a_k}\rangle \neq \langle v_{n_1}^{a_1} \rangle\dots \langle v_{n_k}^{a_k}\rangle$. This issue was resolved in the recent work of Ref.~\cite{Bilandzic:2020csw}, which proposed a new estimator called the Gaussian Estimator (GE).
In the derivation of the GE, a two-dimensional Gaussian distribution was used to approximate multiharmonic flow fluctuations. This leads to the following expression for the estimation of SPC 
\begin{equation}
    \langle \cos (a_1 n_1 \Psi_{n_1} +\dots+ a_k n_k \Psi_{n_k})\rangle_{\text{GE}} = \sqrt{\frac{\pi}{4}}\frac{\langle v_{n_1}^{a_1}\dots v_{n_k}^{a_k} \cos(a_1 n_1 \Psi_{n_1} +\dots+ a_k n_k \Psi_{n_k})\rangle}{\sqrt{ \langle v_{n_1}^{2a_1} \dots  v_{n_k}^{2a_k}\rangle}}\,.
    \label{eq. GE}
\end{equation}
Again, both the numerator and denominator on the right-hand side of Eq.~\eqref{eq. GE} can be estimated using Eq.~\eqref{eq:generalResult}. In contrast to the SP method (Eq.~\eqref{eq. SP-Method}), the denominator of the GE in Eq.~\eqref{eq. GE} contains only the multivariate joint moment of the flow amplitudes $\langle v_{n_1}^{2a_1} \dots v_{n_k}^{2a_k}\rangle$. As such, the GE accounts for the correlations among the flow amplitudes, which overcomes the bias present in the SP method as demonstrated in Ref.~\cite{Bilandzic:2020csw}. Recently, the GE was used by ALICE in the study of SPC in Pb--Pb collisions at a centre-of-mass energy per nucleon pair \snn~$=2.76$~TeV~\cite{ALICE:2023wdn}. The correlations estimated with the GE were significantly smaller compared to the results obtained with the SP method by ATLAS~\cite{ATLAS:2014ndd} and ALICE~\cite{ALICE:2017fcd}. This observation is qualitatively in agreement with the results in Ref.~\cite{Bilandzic:2020csw} and was attributed to the minimisation of the bias from neglecting the correlations among flow amplitudes. The removal of the bias is important in order to extract independent information about symmetry planes.

In this article, the GE is utilised in Pb--Pb collisions at \snn~$=5.02$~TeV. The large data set allows more SPC combinations to be studied including, for the first time, measurements involving up to five different symmetry planes. These provide a detailed study of how symmetry planes of different harmonics interplay, which is crucial for the understanding of the impact of linear and non-linear hydrodynamic responses of the medium, whose evolution is governed by the strong force as described by the quantum chromodynamics (QCD). Additionally, by comparing results with those from Ref.~\cite{ALICE:2023wdn}, the energy dependence of SPC is explored.

The article is structured as follows. Section~\ref{sec:eda} presents the ALICE detector, the event and track selections of the analysed data set, and the estimation of the statistical and systematic uncertainties. The results and their comparisons with state-of-the-art theoretical predictions are discussed in Sec.~\ref{sec:res}. Section~\ref{sec:summary} provides a summary of the article. Additional results are shown in Appendices~\ref{sec:app1}, ~\ref{sec:app2}, and~\ref{sec:app3}.

%============================================================%
\section{Data analysis}
\label{sec:eda}

This analysis is conducted using the Pb--Pb collision data sets at \fivenn recorded by ALICE in 2015 and 2018. The detailed descriptions of the various detectors and their performance are given in Refs.~\cite{ALICE:2008ngc, ALICE:2014sbx}.

For the triggering, event selection, and centrality determination, the V0 detector is used. This detector consists of two scintillator arrays, V0A and V0C~\cite{ALICE:2008ngc, ALICE:2013axi}. 
The arrays cover the full azimuth and extend in pseudorapidity ($\eta$) over the ranges $2.8 < \eta < 5.1$ (V0A) and $-3.7 < \eta < -1.7$ (V0C).
The centrality is determined from the V0 signal and is generally presented as a percentile of the total hadronic Pb--Pb cross section. The most central (head-on) collisions, which produce the largest number of particles, correspond to 0\% percentile, while the more peripheral collisions are denoted with higher percentiles.
Charged particle tracks are reconstructed using the Time Projection Chamber (TPC)~\cite{Alme:2010ke}, which covers the full azimuthal angle and $|\eta| < 0.9$ in the longitudinal direction. In addition to the TPC, the Inner Tracking System (ITS)~\cite{ALICE:2013nwm, ALICE:2010tia} is used for track reconstruction. The ITS consists of six silicon layers, where the two innermost layers constitute the Silicon Pixel Detector (SPD).
The SPD provides high-resolution space points for the determination of the track parameters in the vicinity of the beam axis and for the primary and secondary vertex reconstruction. Both the ITS and TPC, are placed in a homogeneous solenoidal magnetic field of 0.5~T in beam direction, whose polarity was reversed during the data taking.

For both 2015 and 2018 data sets, events are selected with the minimum bias (MB) trigger, which requires a coincident trigger signal from both V0A and V0C. Additional triggers for central and semicentral collisions are used for 2018 data to include more events in the centrality percentiles 0--10\% and 30--50\%, respectively. Furthermore, for the data taken in 2018, centrality flattening is applied by rejecting events to ensure a uniform distribution per centrality interval. With the uniform distribution, the centrality interval becomes properly averaged. Overall seven centrality classes within the range of $0-60$\% have been used for the analysis.

For the analysis, triggered events with a reconstructed primary vertex within $\pm8.0$~cm from the nominal interaction point along the beam line are selected to ensure a full geometrical acceptance of the ITS for $|\eta| < 0.9$. 
Background events such as beam--gas collisions and pileup were removed using the timing information from the V0 detector, the correlation between the primary vertex positions reconstructed from SPD tracklets and ITS+TPC tracks, and the correlation between the number of hits in the TPC and in the outer layers of the ITS~\cite{Arslandok:2022dyb}.
A further rejection of particles produced in out-of-bunch pileup collisions was achieved by selecting tracks that either have a hit in the SPD, which has short readout time of 300 ns, or have a production time, measured with the Time-Of-Flight detector~\cite{TOF}, compatible with the bunch crossing that fired the trigger.
Furthermore, this analysis requires correlators involving up to 14 particles; therefore, a minimum of 14 reconstructed tracks per event was necessary to capture all needed particle correlations. Following these event selection criteria, the number of events used in the analysis is approximately 210 million. 

The pseudorapidity and transverse momentum of the tracks are required to be in in the intervals $|\eta| < 0.8$ and $0.2 < \pt < 5.0$~GeV/\textit{c}. The lower \pt limit is set to remove tracks with low reconstruction efficiency, while the upper limit is set to reduce contamination from jets, which generally dominate at higher $\pt$.
For each track, the distance of closest approach (DCA) to the primary vertex is required to be less than 2~cm along the longitudinal direction and less than $0.0105 + 0.0350/\pt^{1.1}$~cm in the transverse plane, where the $\pt$~is in GeV/\textit{c}. Thus, in the transverse direction, the DCA ranges from 0.016~cm (\pt = 5.0~GeV/\textit{c}) to 0.22~cm (\pt = 0.2~GeV/\textit{c}).
The DCA value requirements on the tracks are used to reduce contamination from secondary particles originating from weak decays and interactions with the detector material. The contamination is largest for particles with low-\pt and it ranges from 20\% (10\%) for $\pt<0.1$~GeV/$c$ to less than 2\% (1\%) for $\pt>1.0$~GeV/$c$ in central (peripheral) collisions~\cite{ALICE:2018vuu}. The contributions of correlations that do not arise from anisotropic flow were tested with HIJING simulations. The check was conducted separately for the numerator and denominator for all SPC. The resulting correlators from HIJING were compatible with zero, and thus having no contributions from non-flow.  
Furthermore, only tracks with a minimum of 70 out of 159 TPC space points are selected to guarantee precise tracking. The $\chi^2$ per space point value is required to be between 0.1 and 2.5 to ensure a good fit of the tracks. In addition, a minimum of 2 hits in the ITS is required for each track. These selection criteria are consistent with those in Refs.\cite{ALICE:2022zks, ALICE:2023lwx}. 

The reconstructed tracks are corrected for their non-uniform reconstruction efficiency (NUE) as a function of $\pt$ and for non-uniform azimuthal acceptance (NUA). The NUE corrections are obtained by constructing $\pt$-dependent weights for the tracks from simulations performed with HIJING (Heavy-Ion Jet INteraction Generator)~\cite{Gyulassy:1994ew} event generator in combination with GEANT~3~\cite{Brun:1994aa} transport code including a detailed description of the ALICE detectors and their efficiency. The NUE corrections are used to account for the reconstruction efficiency as a function of \pt and the contamination from secondary particles~\cite{ALICE:2018vuu}.
The applied NUA correction is data-driven such that the data are first scanned to obtain weights to each cell in longitudinal ($\eta$) and azimuthal ($\varphi$) directions for each value of the longitudinal position of the primary vertex. The weights are then applied for each particle in the analysis.

Statistical uncertainties were determined by measuring variances of the numerator and denominator in Eq.~\eqref{eq. GE} separately and then evaluating the full uncertainty with standard propagation of uncertainty. The procedure is identical to the one used in Ref.~\cite{ALICE:2023wdn} for the analysis of Pb--Pb collisions at \twosevensixnn. 

To evaluate the systematic uncertainties, the selection criteria are varied one by one and the results are compared with those obtained with the nominal selections. The recommendations from Ref.~\cite{Barlow:2002yb} are used to determine the significance of separate systematic trials. This test for statistical significance is sometimes called \textit{the Barlow test}. If for a trial the difference to the default value divided by the uncertainty of that difference is greater than one, then the relative variation of the results obtained with this modified selection is considered to be significant within statistical uncertainties. The trial is then included as a systematic uncertainty. Finally, the total systematic uncertainty is calculated for each centrality class by summing the significant sources in quadrature.

In the following, the different specific contributions to the systematic uncertainty are presented and for each of them the average value of the uncertainty (averaged over centrality classes and SPC observables) is reported in parentheses.
The impact of the variable used to determine the centrality is estimated by substituting the default centrality estimator from energy deposition in the V0 scintillator with an estimate based on the number of hits in the first layers of the ITS (12\%).
In the event selection, systematic uncertainties are estimated by varying the selection on the longitudinal position of the primary vertex from $\pm 8$ to $\pm 7$ and $\pm 9$~cm (9\%), and by using a tighter pileup rejection criteria (7\%) to reduce the contamination of events from different bunch crossings~\cite{Arslandok:2022dyb}. Additionally, the effect of magnetic field polarity of the ALICE's solenoid magnet is examined by repeating the analysis for both field polarities separately (13\%). By default, both field polarities are combined in the final results.  
To evaluate systematic uncertainties in the track reconstruction, the selection on the minimum number of TPC space points is changed from 70 to 80 (8\%), and the required $\chi^2$ value per TPC cluster of the reconstruction fit is changed from [0.1, 2.5] to [0.1, 2.3] (7\%).
Furthermore, the applied threshold on the longitudinal DCA value of the reconstructed tracks is changed from 2 to 1~cm (7\%).
Finally, the default tracking selection is changed to another set of track-selection criteria, where the different conditions are set on the information from the ITS and TPC to yield flat azimuthal particle distribution (19\%).

%============================================================%
\section{Results}
\label{sec:res}
This section presents and discusses the SPC results. Section~\ref{subsec:magn} details the results as a function of centrality for all studied observables. Section~\ref{subsec:independence} explores the interdependence among the SPC observables, while in Sec.~\ref{subsec:hydro}, the SPC are compared with predictions from hydrodynamic models. Finally, Sec.~\ref{subsec:sNN_main} compares the measurements with results from Pb–Pb collisions at \twosevensixnn~\cite{ALICE:2023wdn}. Additional figures are provided in Appendices~\ref{sec:app1} and~\ref{sec:app3}.

\subsection{Comparison of magnitudes among various SPC}
\label{subsec:magn}
\begin{figure}[]
    \centering
    \includegraphics[width=\textwidth]{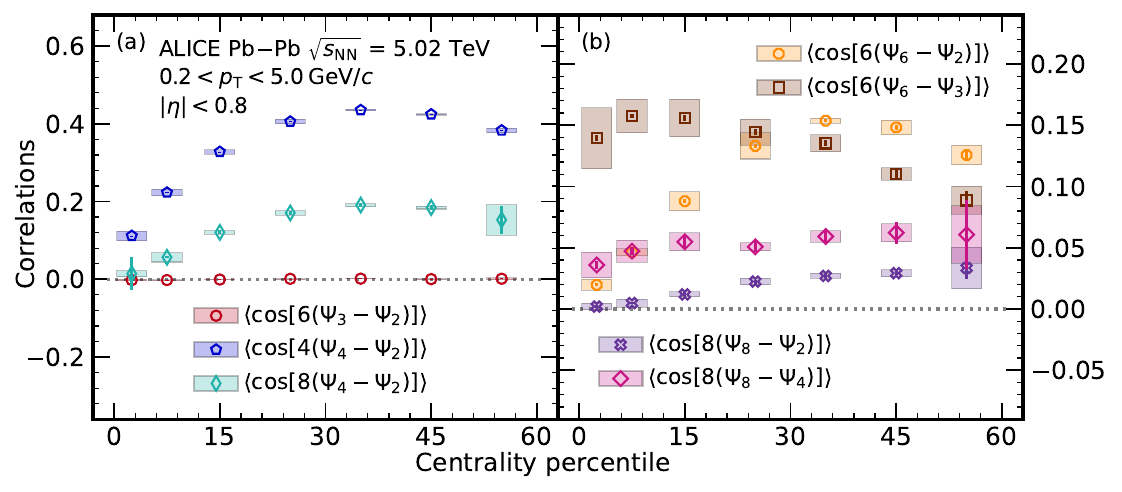}
    \caption{Magnitude of SPC between two symmetry planes. The statistical and systematic uncertainties are depicted as lines and boxes, respectively. }
    \label{fig:2spc_magn}
\end{figure}

Figure~\ref{fig:2spc_magn} shows the two-harmonic SPC magnitudes as a function of centrality.
The strongest magnitudes are obtained for $\langle\cos[4(\Psi_4-\Psi_2)]\rangle$ in all centrality classes, with a clear increasing trend towards semicentral collisions. Other two-harmonic observables showing a similar trend are $\langle\cos[8(\Psi_4-\Psi_2)]\rangle$, $\langle\cos[8(\Psi_8-\Psi_2)]\rangle$, and $\langle\cos[6(\Psi_6-\Psi_2)]\rangle$. Whereas $\langle\cos[8(\Psi_4-\Psi_2)]\rangle$, $\langle\cos[4(\Psi_4-\Psi_2)]\rangle$, and $\langle\cos[6(\Psi_6-\Psi_2)]\rangle$ have slight decrease in trend at the most peripheral collisions, $\langle\cos[8(\Psi_8-\Psi_2)]\rangle$ seems to saturate for centrality percentiles above 30\%. 
For the aforementioned observables, the correlations come from the initial geometry and the non-linear response of the medium. In central collisions, the correlations are weaker due to the more isotropic collision geometry.
In semicentral collisions, the initial anisotropy causes stronger correlations, which are carried to the final state.
The comparison with initial-state calculations are discussed in Sec.~\ref{subsec:hydro}.
The SPC $\langle\cos[6(\Psi_3-\Psi_2)]\rangle$ and $\langle\cos[8(\Psi_8-\Psi_4)]\rangle$ show no centrality dependence, with $\langle\cos[6(\Psi_3-\Psi_2)]\rangle$ being compatible with zero with $n_\sigma=2.07$ significance. The reported statistical significance, represented by the number $n_\sigma$ of Gaussian standard deviations, is obtained from the p-value of the $\chi^2$ distribution. A strong magnitude in central and semicentral collisions with a decreasing trend towards peripheral collisions can be seen for $\langle\cos[6(\Psi_6-\Psi_3)]\rangle$. The trend is slightly different from the other observables and a possible explanation is discussed below. 

In Ref.~\cite{ALICE:2023wdn}, it was noted that the magnitudes are associated with the number of correlating particles. The SPC measured with fewer-particle correlators have larger magnitudes.
Similar results can be seen here, with a slight exception of the three-particle correlator $\langle \cos[8(\Psi_8-\Psi_4)] \rangle$, for which the correlation's lower magnitude is due to high-order symmetry plane, $\Psi_8$. 
In addition, the magnitudes of SPC decrease both with the number of correlating particles and at higher orders. Triangular and higher-order flow in Pb--Pb collisions emerges from the granular distribution of colliding nucleons. These higher-ordered symmetries weaken due to the limited number of participating nucleons in the collision. This means that it becomes less likely for higher-order regular shapes, e.g. a shape with eight-leaves, to develop persistently from a collision to another. Consequently, these symmetry planes become more difficult to detect and they exhibit weaker correlation strength.
The number of correlated particles for a SPC $\langle \cos (a_1 n_1 \Psi_{n_1} +\dots+ a_k n_k \Psi_{n_k})\rangle$ is $\sum a_i$ (note that $a_i>0$). 
Furthermore, the reason for smaller magnitudes of correlators with fewer correlating particles could come from the fluctuations of flow-vector distributions. The particle correlators can be linked to the distribution of the flow-vector fluctuations~\cite{Mehrabpour:2018kjs,Yan:2014afa}, which result from the granular distribution of nucleons inside the colliding nuclei. According to the central limit theorem (CLT), the sample average of independent variables, i.e.~the position of the nucleons in the overlapping region, approaches a Gaussian distribution. This was also noted in the SPC analysis at lower energies~\cite{ALICE:2023wdn}. For distributions close to a Gaussian distribution, the higher-order moments (skewness, kurtosis, etc.) are weaker, and hence, the higher-order correlators are weaker. In central collisions, the distributions tend to be more Gaussian than in peripheral collisions, which can be reflected in the ordering of the correlations in central collisions. The three-particle correlator $\langle\cos[6(\Psi_6-\Psi_3)]\rangle$ is stronger than $\langle\cos[6(\Psi_6-\Psi_2)]\rangle$ in central collisions. For the 20--30\% centrality interval, the ordering begins to swap as $\langle\cos[6(\Psi_6-\Psi_2)]\rangle$ starts to show stronger correlations than $\langle\cos[6(\Psi_6-\Psi_3)]\rangle$. This is due to the increasing deviations from Gaussian, causing higher moments to have more impact. 
Finally, it should be noted that a six-particle correlator $\langle \cos [8(\Psi_4-\Psi_2)] \rangle$ has a large magnitude as it is the second moment of $\langle \cos[4(\Psi_4-\Psi_2)] \rangle$. This will be discussed later in this section.

\begin{figure}[]
    \centering
    \includegraphics[width=\textwidth]{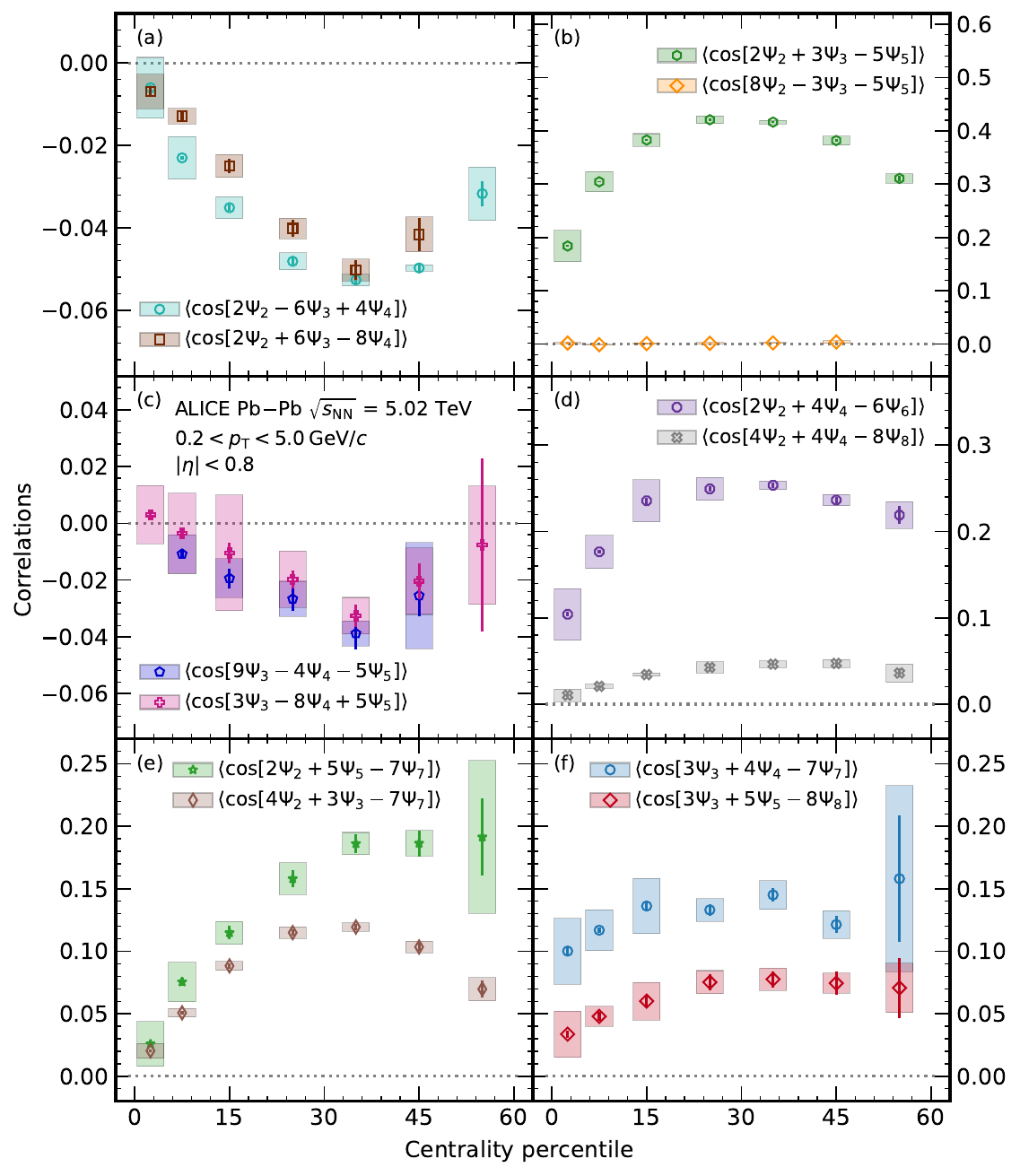}
    \caption{Magnitude of SPC among three symmetry planes. The statistical and systematic uncertainties are depicted as lines and boxes, respectively.}
    \label{fig:3spc_magn}
\end{figure}

The results for three-harmonic SPC as a function of centrality are shown in Fig.~\ref{fig:3spc_magn}.
As in the case of two-harmonic SPC, the three-harmonic SPC also displays the ordering of the magnitudes strongly associated with the number of correlating particles. The largest magnitudes are shown by three-particle correlators, which are listed in order from the strongest to the weakest: $\langle \cos [ 2\Psi_2 + 3\Psi_3 - 5 \Psi_5 ] \rangle$, $\langle \cos [ 2\Psi_2 + 4\Psi_4 - 6 \Psi_6 ] \rangle$, $\langle \cos [ 2\Psi_2 + 5\Psi_5 - 7 \Psi_7 ] \rangle$, $\langle \cos [ 3\Psi_3 + 4\Psi_4 - 7 \Psi_7 ] \rangle$, and $\langle \cos [ 3\Psi_3 + 5\Psi_5 - 8 \Psi_8 ] \rangle$. The magnitude ordering is described by the order of the symmetry planes, such that the weaker magnitudes are shown by the SPC with higher-order symmetry planes.
Correlations among symmetry planes can be intuitively expected to arise from the cumulant expansion of harmonics~\cite{Teaney:2010vd, Teaney:2012ke, Teaney:2013dta}, where higher-order harmonics have non-linear contributions from the lower order harmonics. 
The contributions contain a direct correlation between the participant planes, which can be expected to translate to the final state. This could be the reason why there is a positive correlation among symmetry planes $\Psi_a$, $\Psi_b$, and $\Psi_c$ for which $na+mb=c$, where $n,\,m \in \mathbb{N}$. For example, a cumulant expansion of the fifth-order harmonic has contributions from the second and third-order harmonics, so SPC among $\Psi_2$, $\Psi_3$, and $\Psi_5$, i.e.~$\langle \cos [ 2\Psi_2 + 3\Psi_3 - 5 \Psi_5 ] \rangle$, should have a strong correlation. The four-particle correlators, $\langle \cos [ 4\Psi_2 + 4\Psi_4 - 8 \Psi_8 ] \rangle$ and $\langle \cos [ 4\Psi_2 + 3\Psi_3 - 7 \Psi_7 ] \rangle$, as expected, have smaller magnitudes than the three-particle correlators with similar harmonics.

The observables $\langle \cos[2\Psi_2 - 6\Psi_3 + 4\Psi_4] \rangle$ and $\langle \cos[2\Psi_2 + 6\Psi_3 - 8\Psi_4] \rangle$ show negative correlation among the symmetry planes $\Psi_2$, $\Psi_3$, and $\Psi_4$. Similarly, the symmetry planes $\Psi_3$, $\Psi_4$, and $\Psi_5$ are negatively correlated as shown in Fig.~\ref{fig:3spc_magn}~(c). The SPC analysis at \twosevensixnn also measured a negative correlation in $\langle \cos[2\Psi_2 - 6\Psi_3 + 4\Psi_4] \rangle$, whereas the others were not measured~\cite{ALICE:2023wdn}.
A similar negative correlation was seen among flow harmonics $v_2$, $v_3$, and $v_4$ in the paper that studied event-by-event correlations using higher-order SC in Pb--Pb collisions at \twosevensixnn~\cite{ALICE:2021klf}.
However, the correlation among $v_3$, $v_4$, and $v_5$ was compatible with zero for SC. 
Furthermore, for three-harmonic SPC, the negative correlations are generally smaller in absolute magnitude than the positive correlations.

The lowest magnitude is shown by $\langle \cos[8\Psi_2 - 3\Psi_3 - 5\Psi_5] \rangle$. It is compatible with zero, having only a $n_\sigma = 0.82$ significance of a positive value. 
Although the correlation is minimal, it is slightly positive.
A detailed view of this observable is presented in the model comparison section (Sec.~\ref{subsec:hydro}) in Fig.~\ref{fig:hydro_fig2}. 

\begin{figure}[]
    \centering
    \includegraphics[width=\textwidth]{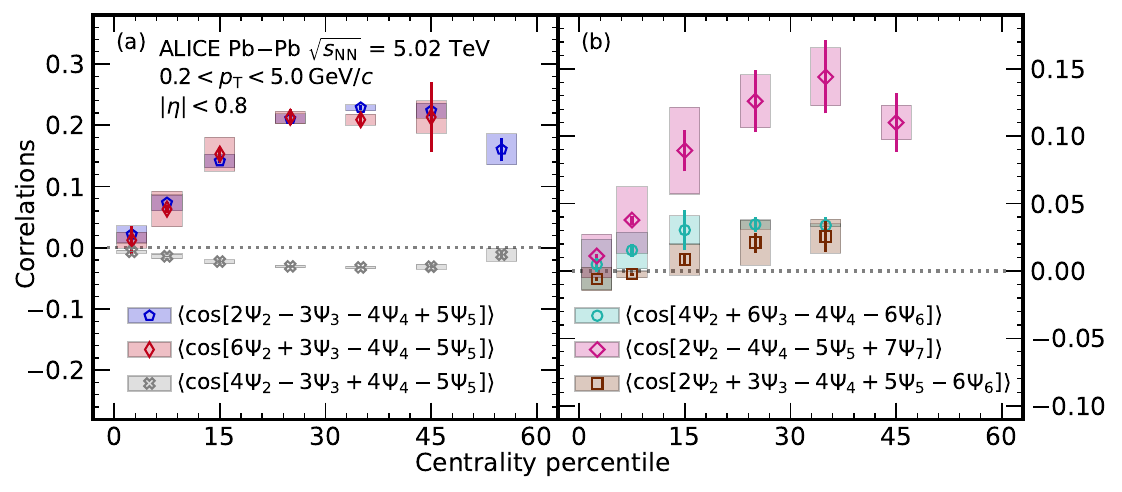}
    \caption{Magnitude of SPC among four and five symmetry planes. The statistical and systematic uncertainties are depicted as lines and boxes, respectively.}
    \label{fig:45spc_magn}
\end{figure}

The magnitudes of four- and the first-ever measured five-harmonic SPC as a function of centrality are displayed in Fig.~\ref{fig:45spc_magn}.
The five-harmonic SPC $\langle \cos[2\Psi_2 + 3\Psi_3 - 4\Psi_4 + 5\Psi_5 - 6\Psi_6] \rangle$ shows a small (if any) increase in magnitude towards semiperipheral collisions, where the correlation is positive within uncertainties. As observed with the two- and three-harmonic SPC, the trends vary among observables containing the same symmetry planes ($\Psi_2,\, \Psi_3,\, \Psi_4,\, \Psi_5$). Specifically, $\langle \cos[2\Psi_2 - 3\Psi_3 - 4\Psi_4 +5\Psi_5] \rangle$ and $\langle \cos[6\Psi_2 + 3\Psi_3 - 4\Psi_4 - 5\Psi_5] \rangle$ have positive magnitudes, while $\langle \cos[4\Psi_2 - 3\Psi_3 + 4\Psi_4 - 5\Psi_5] \rangle$ exhibits slightly negative correlations.
The possible reason for this is discussed in Sec.~\ref{subsec:independence}, as well as the relatively strong correlation of the six-particle correlator $\langle \cos[6\Psi_2 + 3\Psi_3 - 4\Psi_4 - 5\Psi_5] \rangle$. A weak correlation is exhibited by $\langle \cos [4\Psi_2 + 6\Psi_3 - 4\Psi_4 - 6\Psi_6] \rangle$ with a weak centrality dependence. A strong centrality dependence is shown by $\langle \cos [2\Psi_2 - 4\Psi_4 -5\Psi_5 + 7\Psi_7] \rangle$ even within the large uncertainties.

\subsection{Observable interdependence}
\label{subsec:independence}
Many of the SPC observables presented in the previous section are built from different arrangements of the same symmetry planes. Due to the trigonometric properties of the cosine function, some of these results may trivially depend on each other. The starting point to probe the interdependence between two SPC is the following identity,
\begin{equation}\label{eq:trig}
    2 \cos(a)\cos(b) = \cos(a+b) + \cos(a-b). 
\end{equation}
Equation~\eqref{eq:trig} states that if three out of the four cosine terms are known, the fourth term can be determined from the three known terms. 
In an experimental setting, one measures averages over all events and, thus, Eq.~\eqref{eq:trig} becomes
\begin{equation}\label{eq:trig2}
    2 \langle \cos(a)\cos(b) \rangle = \langle \cos(a+b) \rangle + \langle \cos(a-b)\rangle. 
\end{equation}
If no direct correlation between $\cos(a)$ and $\cos(b)$ is assumed, the following approximation can be made
\begin{equation}\label{eq:approx}
    \langle \cos(a) \rangle \langle \cos(b) \rangle \approx \langle \cos(a)\cos(b) \rangle.
\end{equation}
Inserting Eq.~\eqref{eq:trig2} into Eq.~\eqref{eq:approx} then gives
\begin{equation}
    2\langle \cos(a) \rangle \langle \cos(b) \rangle \approx  \langle \cos(a+b) \rangle + \langle \cos(a-b)\rangle. \label{eq:approx2}
\end{equation}

\begin{figure}[]
    \centering
    \includegraphics[width=\textwidth]{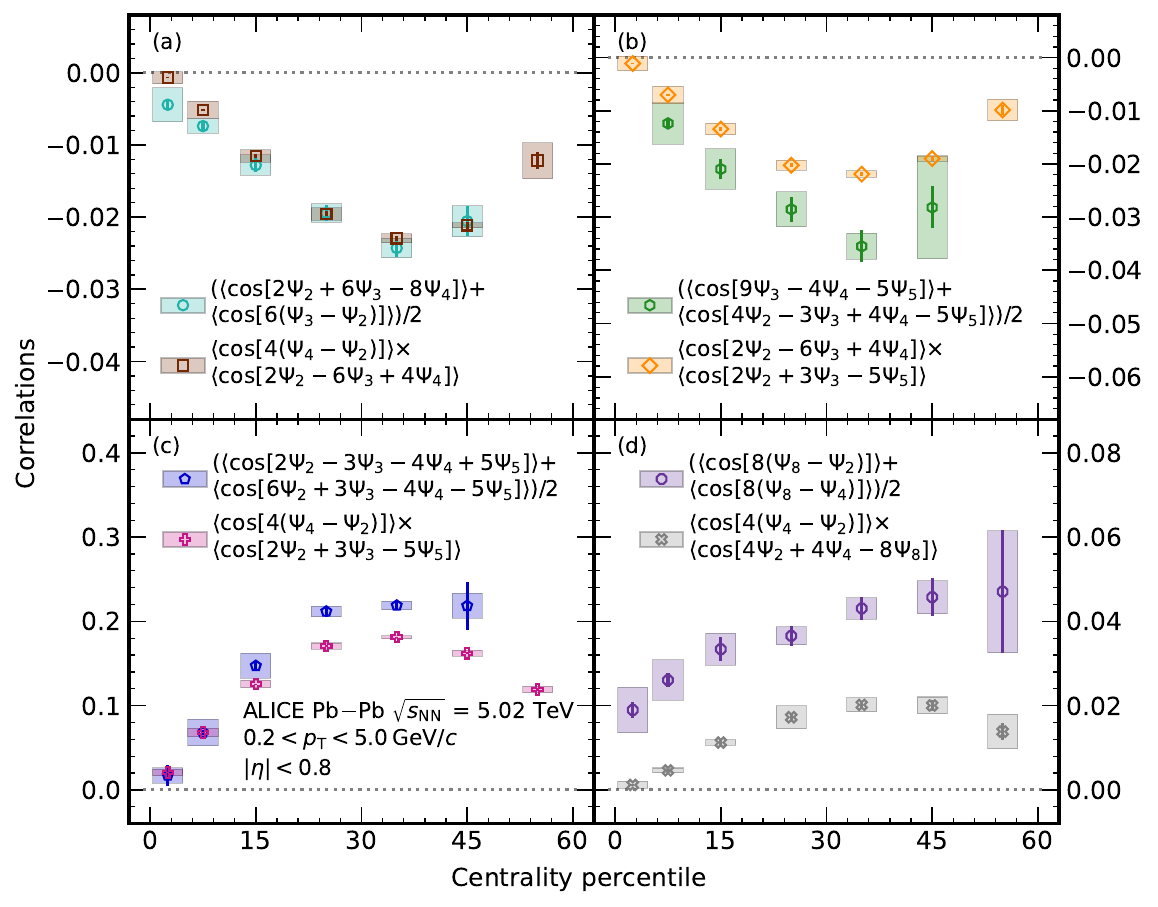}
    \caption{Correlations between different sums of SPC compared with their respective products following Eq.~\eqref{eq:approx2}, in the case of five different SPC. The statistical and systematic uncertainties are depicted as lines and boxes, respectively.}
    \label{fig:decomp}
\end{figure}

Figure~\ref{fig:decomp} presents the interdependence among five different combinations of harmonics through the correlations among different sums of SPC compared with their respective products.
Panel (a) in Fig.~\ref{fig:decomp} shows that the observed dependencies align well with the approximation given by Eq.~\eqref{eq:approx2}, while panels~(b),~(c), and~(d) deviate from it. 
For panel~(b), the $\langle \cos[2\Psi_2 - 6\Psi_3 + 4\Psi_4] \rangle$ and $ \langle \cos[2\Psi_2 + 3\Psi_3 - 5\Psi_5] \rangle$ do not fully decorrelate in the form of Eq.~\eqref{eq:approx}, indicating a potential dependence or correlation between them. 
Since the approximation in Eq.~\eqref{eq:approx2} does not fully apply, this dependence does not fully explain the unexpected negative correlation observed for $\langle \cos[4\Psi_2 - 3\Psi_3 + 4\Psi_4 - 5\Psi_5] \rangle$, as discussed in Sec.~\ref{subsec:magn}. Nonetheless, the negative correlation could then result from the correlation between $\langle \cos [2\Psi_2 - 6\Psi_3 + 4\Psi_4] \rangle $ and $\langle \cos[2\Psi_2 + 3\Psi_3 - 5\Psi_5] \rangle$.
In Sec.~\ref{subsec:magn} it was also noted that $\langle \cos[6\Psi_2 + 3\Psi_3 - 4\Psi_4 - 5\Psi_5] \rangle$ has a relatively large correlation. If the approximation in Eq.~\eqref{eq:approx2} could be applied with $\langle \cos[6\Psi_2 + 3\Psi_3 - 4\Psi_4 - 5\Psi_5] \rangle$, then the large correlation could be explained directly 
with the interdependence of the other three harmonics found in the appliance of the approximation.
However, it can be seen in panel~(c) that the observables only follow the approximation from Eq.~\eqref{eq:approx2} in central collisions, but start disagreeing for centrality percentiles above 20\%. As they do not agree in all centrality intervals, the cause for the large correlation cannot be explained fully by the interdependence. 
Finally in panel~(d), the difference between left- and right-hand sides of Eq.~\eqref{eq:approx2} results from a correlation between $\langle \cos [4(\Psi_4 - \Psi_2)] \rangle$ and $\langle \cos [4\Psi_4 + 4\Psi_2 - 8\Psi_8] \rangle$.

To draw comprehensive conclusions, it is essential to quantify the correlations between different SPC observables. One potential approach is to measure the cumulants of symmetry plane correlations (CSC) defined in Ref.~\cite{Bilandzic:2021rgb}. Nevertheless, the CSC in Ref.~\cite{Bilandzic:2021rgb} pose strong requirements of correlating SPC to not share any symmetry planes, $\Psi_n$. For example, if SPC A is composed of symmetry planes $\Psi_a$ and $\Psi_b$, and SPC B of $\Psi_c$ and $\Psi_d$, then for CSC between SPC A and B it is required that $a, b \neq c, d$. These requirements are not satisfied with most observables presented in this subsection. Future studies are needed to address these correlations between SPC in-depth and check for less strict requirements in the CSC definition.

\subsection{Comparison with models}
\label{subsec:hydro}
These new experimental measurements of SPC are compared to  predictions from two different state-of-the-art hydrodynamic models.
The first model is the event-by-event EKRT+hydrodynamics~\cite{Niemi:2015qia, Hirvonen:2022xfv}, which combines next-to-leading order perturbative QCD and a saturation model~\cite{Paatelainen:2012at, Paatelainen:2013eea} to determine the initial energy density profiles. The evolution of the QGP medium is described with viscous hydrodynamics, with a parameterisation of the specific shear viscosity, $\eta/s(T) = \mathrm{dyn}$, based on dynamical freeze-out conditions~\cite{Hirvonen:2022xfv}. It is important to note that this model does not include a hadronic afterburner, instead the hadronic interactions and transport processes are carried out in the hydrodynamic framework that includes partial chemical freeze-out and dynamical kinetic freeze-out condition.. As such, the final-state predictions are derived directly with particles obtained at the hydrodynamic surface. The effect of hadronic interactions on SPC are discussed in detail in Appendix~\ref{sec:app2}. The calculations presented in this paper were obtained with $10^5$ events.
The second model is \trento+VISH(2+1)+UrQMD~\cite{Bass:1998ca, Bleicher:1999xi, Song:2007ux, Shen:2014vra}, referred to hereafter as \trento+iEBE-VISHNU+UrQMD.
The \trento~model~\cite{Moreland:2014oya} is an initial-state model introduced to study high-energy nuclear collisions from pp to A--A systems. It is based on the use of a generalised mean of the nuclei's energy density, with a free parameter whose value allows \trento~to reproduce the behaviour of different models for initial conditions.
Its output is then connected to the causal hydrodynamic evolution model, VISH(2+1)~\cite{Shen:2014vra}. The hadronisation and evolution of the final-state particles are described with the UrQMD model~\cite{Bass:1998ca, Bleicher:1999xi}.
The free parameters of this hybrid model, like the initial conditions or QGP transport properties, were determined using the global Bayesian analysis~\cite{Parkkila:2021yha}. About 198 million events were simulated to extract the predictions presented in this paper. The initial-state calculations from \trento~were obtained using one million generated events.

In heavy-ion collisions, the anisotropies in the final-state particle azimuthal distribution originate from the medium response to the anisotropies in the initial-state geometry.
As in the final state, the asymmetries in the initial geometry can be described with a Fourier series, where the flow amplitudes $v_n$ are replaced by the moments of the initial energy density, the eccentricities $\epsilon_n$, and the symmetry planes $\Psi_n$ by the participant planes $\Phi_n$~\cite{Gardim:2011xv}.
Another description of the initial geometry relies on the cumulants of the initial energy density. It was shown in Refs.~\cite{Teaney:2012ke, Teaney:2013dta, ALICE:2023wdn} that the cumulant formulation leads to a more accurate representation of the asymmetric geometry. This is due to lower-order contributions in the higher-order eccentricities that are removed with the cumulant approach. As such, only the cumulant expansion is used in this study to extract the initial-state predictions. As an example, a cumulant expansion of fourth-order anisotropy can be expressed as
\begin{equation}\label{eq:nonlinear}
    c_4e^{i4\Phi_4} = \varepsilon_4 e^{i4\phi_{\,4}} + 3 \Big(\frac{\{r^2\}^2}{\{r^4\}} \Big) \varepsilon_2^2e^{i4\phi_{\,2}},
\end{equation}
where $\varepsilon$ describes the eccentricity of the initial energy density distribution and $\{\dots\}$ defines an average with respect to the energy density. In Eq.~\eqref{eq:nonlinear} $\Phi_n$ and $\phi_n$ are defined as participant planes in cumulant and moment expansions, respectively. For the lowest-order harmonics $(n<4)$ these are the same, while the higher-order cumulants start to differ from the moments due to their dependence on the lower-order eccentricities. Detailed descriptions of the cumulant and moment expansions can be found in Refs.~\cite{Teaney:2012ke, Teaney:2013dta}. In the following results, the initial-state predictions are obtained by evaluating correlations between participant planes, $\Phi_n$, defined in the cumulant expansion.   
Due to the difference in magnitudes between initial- and final-state fluctuations, observables like the SC must be normalised to allow for proper comparisons between the different stages of the collisions. This procedure is not necessary for SPC, as the fluctuations of the flow amplitudes are already cancelled out by the ratio in Eq.~\eqref{eq. GE}.
Therefore, the correlations between participant planes and between symmetry planes of the same harmonics can straightforwardly be compared, allowing one to infer on the impact of the medium response on the initial-state correlations.

\begin{figure}[t!]
    \centering
    \includegraphics[width=\textwidth]{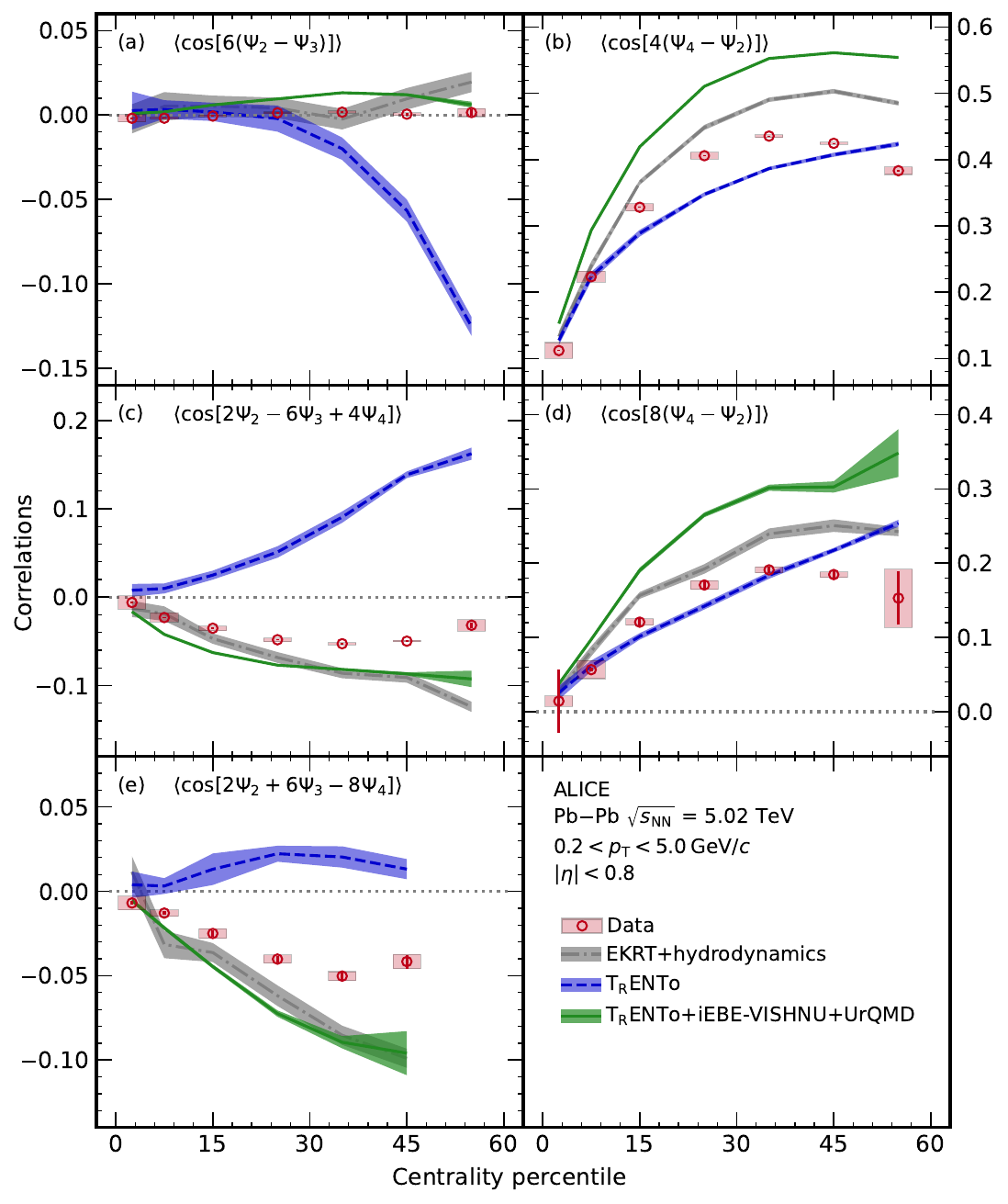}
    \caption{Comparison of the centrality dependence of the correlations involving $\Psi_2$, $\Psi_3$, and $\Psi_4$ (red circles) with the theoretical predictions from \trento+iEBE-VISHNU+UrQMD~\cite{Bass:1998ca, Bleicher:1999xi, Song:2007ux, Shen:2014vra} and EKRT+hydrodynamics~\cite{Hirvonen:2022xfv} shown as green and gray bands, respectively. Initial-state predictions from \trento~calculated with cumulant expansions are shown as blue bands. The lines (boxes) represent the statistical (systematic) uncertainties of the experimental data. The widths of the bands denote the statistical uncertainty of the model predictions.}
    \label{fig:hydro_fig1}
\end{figure}

Figure~\ref{fig:hydro_fig1} presents the correlations between two and three different symmetry planes involving $\Psi_2$, $\Psi_3$, and $\Psi_4$, compared with predictions from both models.
The absence of correlations between $\Psi_2$ and $\Psi_3$ (Fig.~\ref{fig:hydro_fig1}~(a)) seen in the data is well reproduced by EKRT+hydrodynamics within uncertainties. The predictions from \trento+iEBE-VISHNU+UrQMD also capture the flat centrality dependence of $\langle\cos[6(\Psi_3-\Psi_2)]\rangle$ but slightly overestimate the values. The latter confirms the outcome of the comparisons at lower energies discussed in Ref.~\cite{ALICE:2023wdn}. In contrast, the initial-state predictions show a clear deviation from zero for centrality percentiles above 30\%, indicating effects beyond linear hydrodynamic response for $\Psi_2$ and $\Psi_3$. 
For the SPC between $\Psi_2$ and $\Psi_4$ in panels~(b) and~(d), both final-state calculations capture the trend but not the magnitude of the data. The calculations from EKRT+hydrodynamics generally present less discrepancy with the experimental values, and both models show the best agreement with data in central collisions.
As the linear response between $\Psi_2$ and $\Psi_4$ dominates for centralities up to 10\%~\cite{Taghavi:2020gcy}, the difference between data and model calculations in semicentral collisions may originate from the non-linear coupling itself.
Furthermore, the difference in magnitude between $\langle\cos[4(\Psi_4-\Psi_2)]\rangle$ and $\langle\cos[8(\Psi_4-\Psi_2)]\rangle$ is approximately reproduced by both models. Whereas the final-state predictions start to saturate, and even decrease, in the most peripheral collisions considered in this study, the initial-state predictions increase monotonically in the studied centrality range. 
Panels~(c) and~(e) show two different correlations of the three symmetry planes. Both models provide similar predictions and qualitatively reproduce the trend of the experimental data. As observed in the various two-harmonic SPC, data and models are in agreement within their respective uncertainties in central collisions up to 20\%.
The initial-state predictions (obtained with \trento) are positive for both SPC, albeit with different centrality dependence, as opposed to the negative correlation observed for the final-state particles. Such behaviour was already observed in the measurements by ALICE at \twosevensixnn~\cite{ALICE:2023wdn}, where the differences between the initial and final states were attributed to the hydrodynamic evolution of the system.

\begin{figure}[b]
    \centering
    \includegraphics[width=\textwidth]{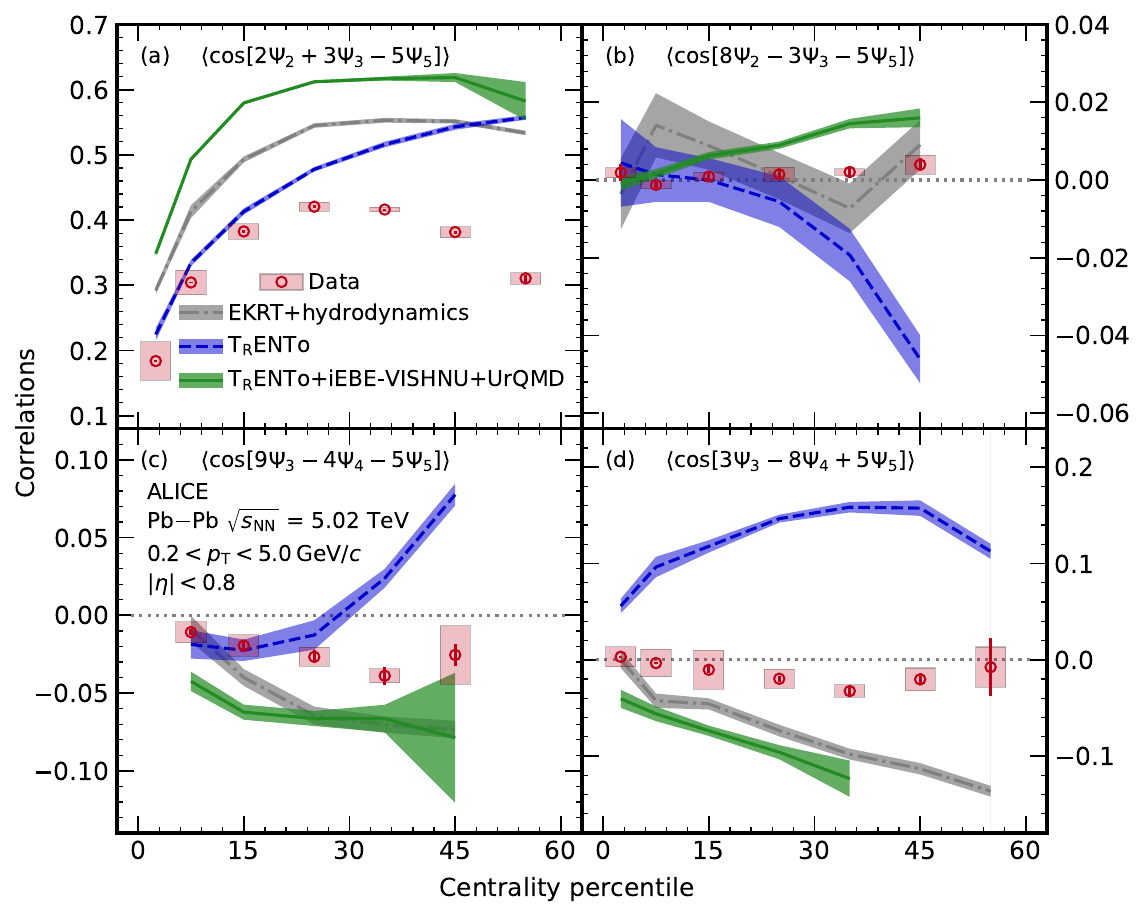}
    \caption{Comparison of the centrality dependence of the three-harmonic correlations involving symmetry planes up to $\Psi_5$ (red circles) with the theoretical predictions from \trento+iEBE-VISHNU+UrQMD~\cite{Bass:1998ca, Bleicher:1999xi, Song:2007ux, Shen:2014vra} and EKRT+hydrodynamics~\cite{Hirvonen:2022xfv} shown as green and gray bands, respectively. Initial-state predictions from \trento~calculated with cumulant expansions are shown as blue bands. The lines (boxes) represent the statistical (systematic) uncertainties of the experimental data. The widths of the bands denote the statistical uncertainty of the model predictions.}
    \label{fig:hydro_fig2}
\end{figure}

The correlations between three symmetry planes up to $\Psi_5$ are shown in Fig.~\ref{fig:hydro_fig2}, alongside their respective model calculations.
The top panels present the SPC between $\Psi_2$, $\Psi_3$, and $\Psi_5$.
As in Fig.~\ref{fig:hydro_fig1}, the predictions from EKRT+hydrodynamics better reproduce the data, although, both models drastically overestimate the magnitude of $\langle\cos[2\Psi_2 + 3\Psi_3 -5\Psi_5]\rangle$. For both observables in the top panels, the trends of initial-state predictions are similar with the final-state predictions and the data in central collisions, but have large increases in the peripheral collisions.
Panels~(c) and~(d) in Fig.~\ref{fig:hydro_fig2} show $\langle\cos[9\Psi_3 -4\Psi_4 -5\Psi_5]\rangle$ and $\langle\cos[3\Psi_3 -8\Psi_4 +5\Psi_5]\rangle$, respectively, measured for the first time with the GE method.
Similarly, as for the correlations between $\Psi_2$, $\Psi_3$, and $\Psi_4$, the model calculations qualitatively describe the negative signature of the data.
An interesting point to highlight is the difference in the behaviour of the initial-state predictions from \trento~between the two observables. While $\langle\cos[9\Psi_3 -4\Psi_4 -5\Psi_5]\rangle$ is negative and agrees with the data in central collisions, it strongly increases to positive values in more peripheral events. This is in contrast with the initial-state predictions for $\langle\cos[3\Psi_3 -8\Psi_4 +5\Psi_5]\rangle$ that are positive in all centralities.
Such difference can be explained by the respective cumulant expansions of both SPC, where $\langle\cos[3\Psi_3 -8\Psi_4 +5\Psi_5]\rangle$ contains additional correlations between the second, third, fourth, and fifth order participant planes.

\begin{figure}[t]
    \centering
    \includegraphics[width=\textwidth]{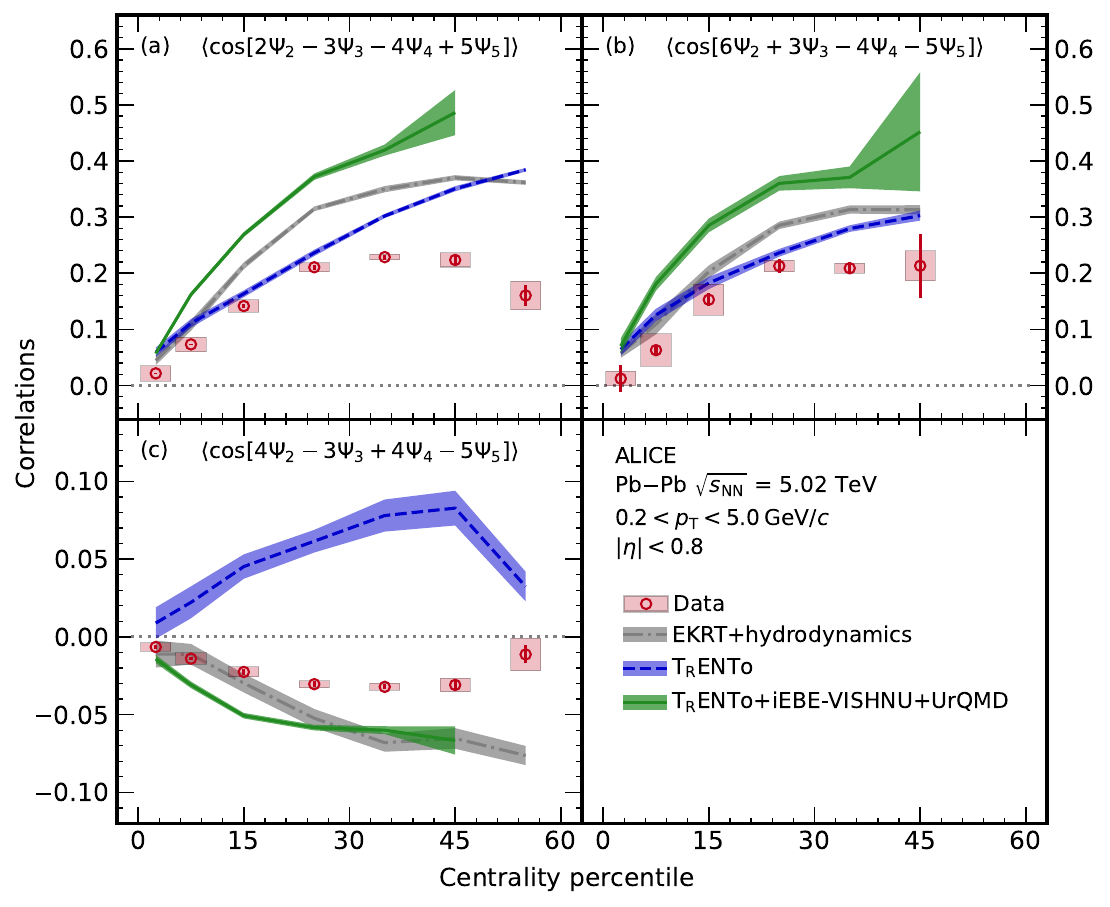}
    \caption{Comparison of the centrality dependence of the correlations between different combinations of $\Psi_2$, $\Psi_3$, $\Psi_4$, and $\Psi_5$ (red circles) with the theoretical predictions from \trento+iEBE-VISHNU+UrQMD~\cite{Bass:1998ca, Bleicher:1999xi, Song:2007ux, Shen:2014vra} and EKRT+hydrodynamics~\cite{Hirvonen:2022xfv} shown as green and gray bands, respectively. Initial-state predictions from \trento~calculated with cumulant expansions are shown as blue bands. The lines (boxes) represent the statistical (systematic) uncertainties of the experimental data. The widths of the bands denote the statistical uncertainty of the model predictions.}
    \label{fig:hydro_fig3}
\end{figure}

Three different correlations between $\Psi_2$, $\Psi_3$, $\Psi_4$, and $\Psi_5$ can be seen in Fig.~\ref{fig:hydro_fig3}.
The SPC $\langle\cos[2\Psi_2 -3\Psi_3 -4\Psi_4 +5\Psi_5]\rangle$ and $\langle\cos[6\Psi_2 +3\Psi_3 -4\Psi_4 -5\Psi_5]\rangle$ shown in panels~(a) and (b), respectively, have similar magnitude and centrality dependence for model predictions and data. Both calculations, \trento+iEBE-VISHNU+UrQMD and EKRT+hydrodynamics, reproduce the centrality dependence of the experimental values but not their magnitude. The predictions from EKRT+hydrodynamics are, as seen previously, closer to the data and in agreement with them within uncertainties for the 0--20\% centrality interval. This range also corresponds to the one where the initial-state calculations from \trento~are in agreement with the data, but not with the final-state results from \trento+iEBE-VISHNU+UrQMD.

Finally, the results for $\langle\cos[4\Psi_2 -3\Psi_3 +4\Psi_4 -5\Psi_5]\rangle$ shown in panel~(c) of Fig.~\ref{fig:hydro_fig3} present the most striking differences with the $\langle\cos[2\Psi_2 -3\Psi_3 -4\Psi_4 +5\Psi_5]\rangle$ and $\langle\cos[6\Psi_2 +3\Psi_3 -4\Psi_4 -5\Psi_5]\rangle$. While the initial-state predictions are still positive, the final-state results for both the models and the experimental measurements are negative. As discussed in Sec.~\ref{subsec:independence}, this observable is expected to be consistent with zero but obtains negative magnitude due to correlations between $\langle \cos [2\Psi_2 - 6\Psi_3 + 4\Psi_4] \rangle $ and $\langle \cos[2\Psi_2 + 3\Psi_3 - 5\Psi_5] \rangle$.

\begin{figure}[t]
    \centering
    \includegraphics[width=\textwidth]{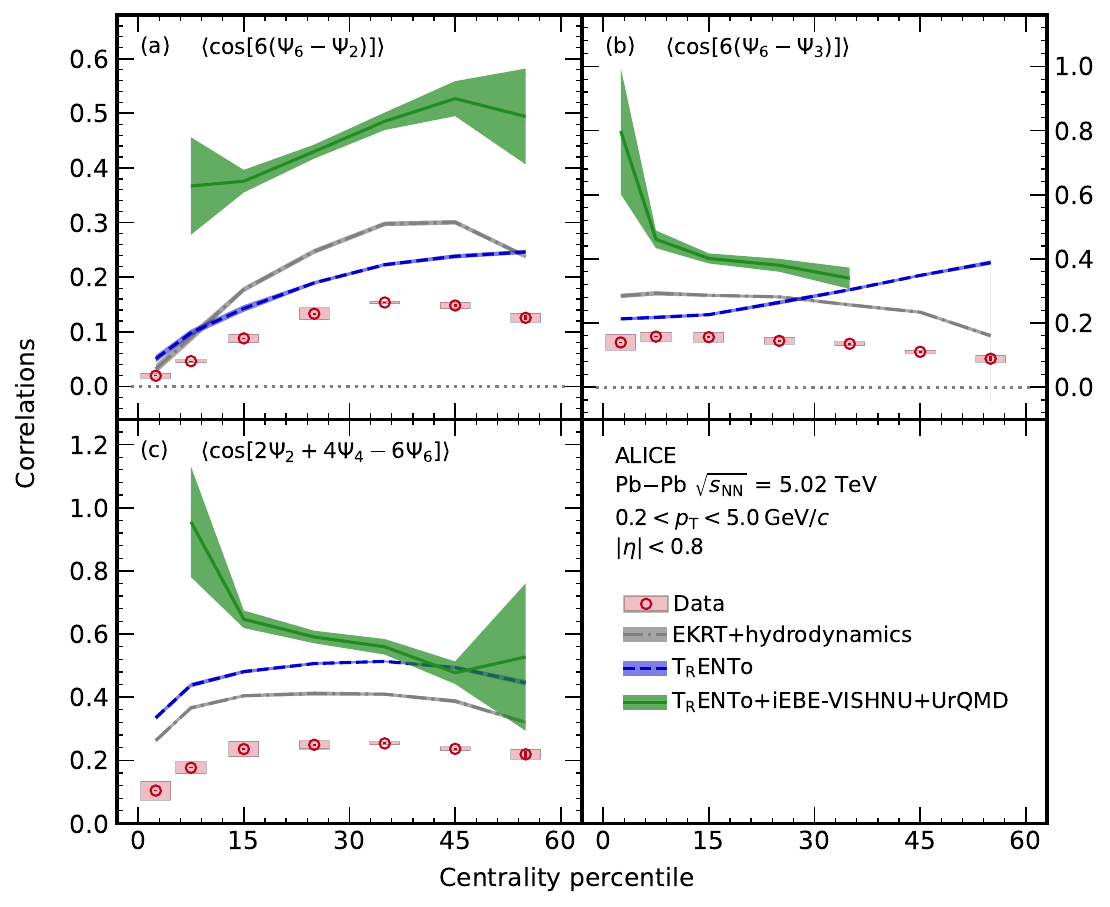}
    \caption{Comparison of the centrality dependence of the correlations involving symmetry planes up to $\Psi_6$ (red circles) with the theoretical predictions from \trento+iEBE-VISHNU+UrQMD~\cite{Bass:1998ca, Bleicher:1999xi, Song:2007ux, Shen:2014vra} and EKRT+hydrodynamics~\cite{Hirvonen:2022xfv} shown as green and gray bands, respectively. Initial-state predictions from \trento~calculated with cumulant expansions are shown as blue bands. The lines (boxes) represent the statistical (systematic) uncertainties of the experimental data. The widths of the bands denote the statistical uncertainty of the model predictions.}
    \label{fig:hydro_fig4}
\end{figure}

Figure~\ref{fig:hydro_fig4} gathers the model predictions for the two- and three-harmonic SPC involving up to $\Psi_6$.
Both final-state models manage to capture the trend but not the magnitude of the data. The initial-state predictions show similar trend as the final-state predictions for $\langle \cos [6(\Psi_6 - \Psi_2)] \rangle $ and $\langle \cos[2\Psi_2 + 4\Psi_4 - 6\Psi_6] \rangle$, while the initial- and final-state predictions for $\langle \cos [6(\Psi_6 - \Psi_3)] \rangle $ have opposite trends. The increasing and decreasing trends of initial- and final-state predictions for $\langle \cos [6(\Psi_6 - \Psi_3)] \rangle $ were also noted in the lower energy analysis in Ref.~\cite{ALICE:2023wdn}.

\subsection{Beam energy dependence}
\label{subsec:sNN_main}
\begin{figure}[t]
    \centering
    \includegraphics[width=\textwidth]{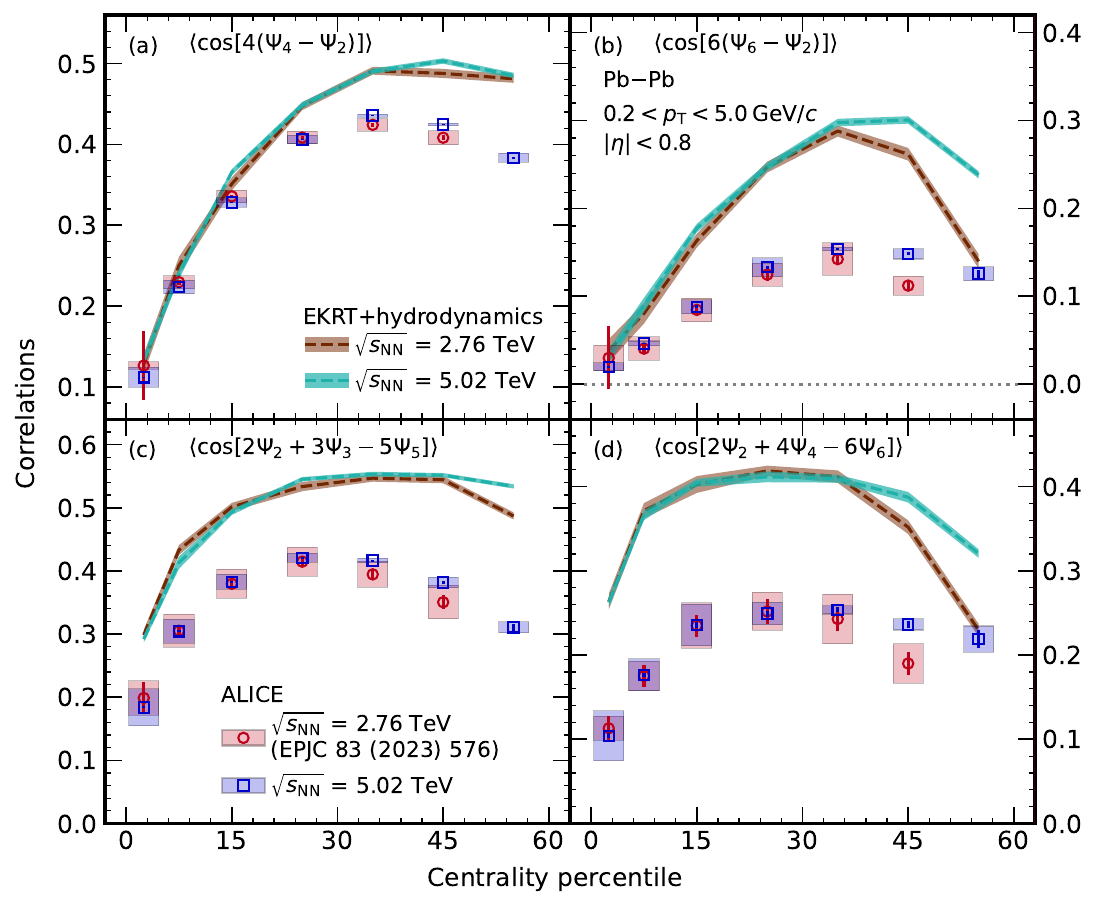}
    \caption{Comparison of different two- and three-harmonic SPC measured at \twosevensixnn (red circles)~\cite{ALICE:2023wdn} and at \fivenn (blue squares) by ALICE. The results are compared with the predictions from EKRT+hydrodynamics~\cite{Hirvonen:2022xfv} at the same centre-of-mass energies (brown and green bands, respectively). The lines (boxes) represent the statistical (systematic) uncertainties in the experimental data. The widths of the bands denote the statistical uncertainty of the model predictions.}
    \label{fig:hydro_sNN}
\end{figure}

Figure~\ref{fig:hydro_sNN} shows different two- and three-harmonic SPC compared with the measurements in Pb--Pb collisions at \twosevensixnn~\cite{ALICE:2023wdn}, along with the EKRT+hydrodynamic model predictions at the two collision energies. No significant energy dependence is observed for the data as the SPC at different beam energies agree with each other within $1.1\sigma$ over the full centrality range (for more details, see Appendix~\ref{sec:app3}).
However, when checking in more detail the evolution with centrality, the  four different combinations of SPC shown in Fig.~\ref{fig:hydro_sNN} present an interesting behaviour for both the experimental data and the calculations from EKRT+hydrodynamics.

In all four cases, the results at both energies are in good agreement for centralities up to 40\%.  Beyond this centrality, the magnitudes of the SPC measured at higher beam energy tend to be slightly larger, even though strong conclusions cannot be drawn considering the current uncertainties. A difference in the same direction and with more pronounced magnitude is seen also in the model predictions. This trend suggests that the longer duration of the QGP phase in Pb--Pb collisions at higher centre-of-mass energy may be a contributing factor in the lesser dissipation of the correlations when compared to the ones at lower beam energies~\cite{Niemi:2015voa, Noronha-Hostler:2015uye}.

%============================================================%
\section{Summary}
\label{sec:summary}
This article presents the measurements of the correlations between different symmetry planes in Pb--Pb collisions at $\sqrt{s_{\mathrm{NN}}} = 5.02$~TeV using the recently developed Gaussian Estimator technique, which allows to eliminate the bias in previously used estimators stemming from correlated flow amplitudes.
The larger data sample allows the analysis to be extended by including more combinations of harmonics than previously reported in Pb--Pb collisions at $\sqrt{s_{\mathrm{NN}}} = 2.76$~TeV~\cite{ALICE:2023wdn}. Furthermore, the first measurement of the correlations between five different symmetry planes is shown.
The investigation of the centrality dependence of the various symmetry plane correlations (SPC) confirms the influence of the order of the SPC and the collision centrality on the correlation strength. In addition, for every harmonic-order SPC, the number of correlating particles has a considerable impact on the hierarchy of the correlation strength as seen previously in the results at $\sqrt{s_{\mathrm{NN}}} = 2.76$~TeV.
No significant energy dependence is observed within the uncertainties in the comparison of the present data with the measurements at the lower centre-of-mass energy.
The measured results are compared with state-of-the-art hydrodynamic model calculations. These calculations qualitatively describe the data showing large differences for correlators exhibiting a significant non-linear response of the medium to initial-state geometry. The EKRT+hydrodynamic model is in better agreement with the data than the \trento+iEBE-VISHNU+UrQMD model, especially for correlations involving $\Psi_2$ and $\Psi_4$. This disagreement between the models can partially be explained with the effects of hadronic afterburner that is included in \trento+iEBE-VISHNU+UrQMD but is absent in EKRT+hydrodynamics. Nevertheless, deviations from the data are observed in both models, mainly for peripheral collisions. 
The correlations between different symmetry planes offer insights into the impact of the medium response on initial-state correlations. The comparisons indicate that the models tend to overestimate the experimental values, possibly due to factors such as the non-linear response from the initial state to the final state or the use of inaccurate parameters.
More specifically, the EKRT and iEBE-VISHNU with {T\raisebox{-.5ex}{R}ENTo} framework exhibit distinct characteristics in their initial-state correlations and subsequent hydrodynamic evolution. These discrepancies provide constraints for refining both initial condition parameters and final-state evolution in heavy-ion collision models. The interplay between the initial conditions and hydrodynamic response, thus, becomes visible in results as deviations from experimental measurements.
The SPC's sensitivity to model parameters not only constrains the QGP parameters, but also helps us understand the underlying physics at each stage of the heavy ion collision and how these stages combine.

In conclusion, this study provides valuable insights into the correlations between symmetry planes in heavy-ion collisions and highlights the need for further improvements in modelling the QCD matter properties in ultrarelativistic heavy-ion collisions.

%%%%%%%%%%%%%%%%%%%%%%%%%%%%%%%%
% end main text 
%%%%%%%%%%%%%%%%%%%%%%%%%%%%%%%%

%%%%% acknowledgements - handled by EB chairs 
\newenvironment{acknowledgement}{\relax}{\relax}
\begin{acknowledgement}
\section*{Acknowledgements}
% add specific acknowledgements here 
The ALICE Collaboration would like to thank Henry Hirvonen for providing the latest predictions from the state-of-the-art hydrodynamic model, event-by-event EKRT+hydrodynamics.
% ...but please don't remove the line below: funding agencies
% will be acknowledged with a custom tex file handled by EB chairs after Collab Round 2
\input{fa_2024-08-10_Opt_C.tex}
\end{acknowledgement}

%%%%%%%% Bibliography 
\bibliographystyle{utphys}   % Remember we use title in the biblio
\bibliography{bibliography}
%\input {bibliography.tex}  

%%%%%%%%%%%%%%%%%%%%%%%%%%%%%%%%
% Appendices: yours (if any) + authorlist
%%%%%%%%%%%%%%%%%%%%%%%%%%%%%%%%
\newpage
\appendix 
\section{Further model comparisons}
\label{sec:app1}
\begin{figure}[h]
    \centering
    \includegraphics[width=\textwidth]{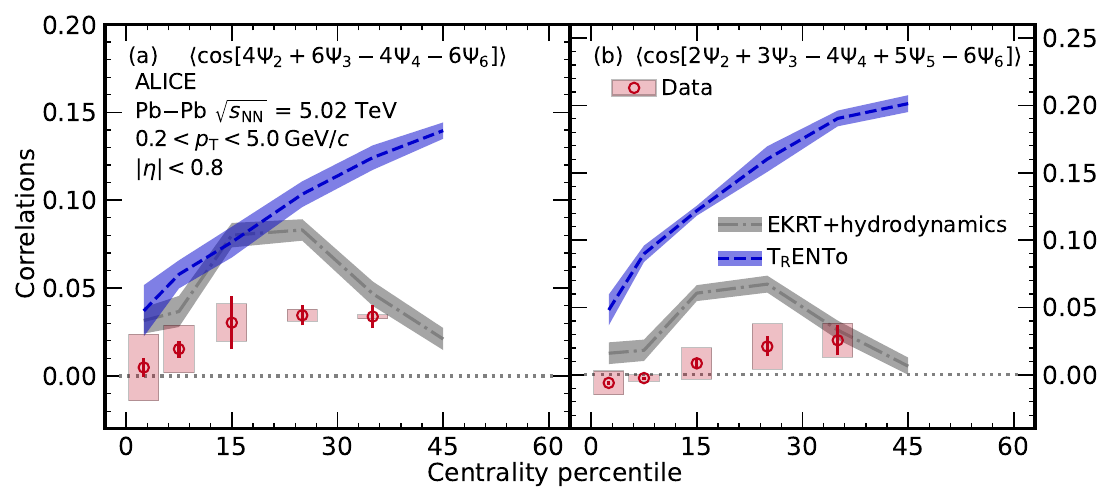}
    \caption{Comparison of the centrality dependence of the correlations among different combinations of $\Psi_2$, $\Psi_3$, $\Psi_4$, $\Psi_5$, and $\Psi_6$ (red circles) with the theoretical predictions from EKRT+hydrodynamics~\cite{Hirvonen:2022xfv} shown in gray bands. Initial-state predictions from \trento~\cite{Bass:1998ca, Bleicher:1999xi, Song:2007ux, Shen:2014vra} calculated with cumulant expansions are shown as blue bands. The lines (boxes) represent the statistical (systematic) uncertainties in the experimental data. The widths of the bands denote the statistical uncertainty of the model predictions.}
    \label{fig:hydro_fig5}
\end{figure}

\begin{figure}[t]
    \centering
    \includegraphics[width=\textwidth]{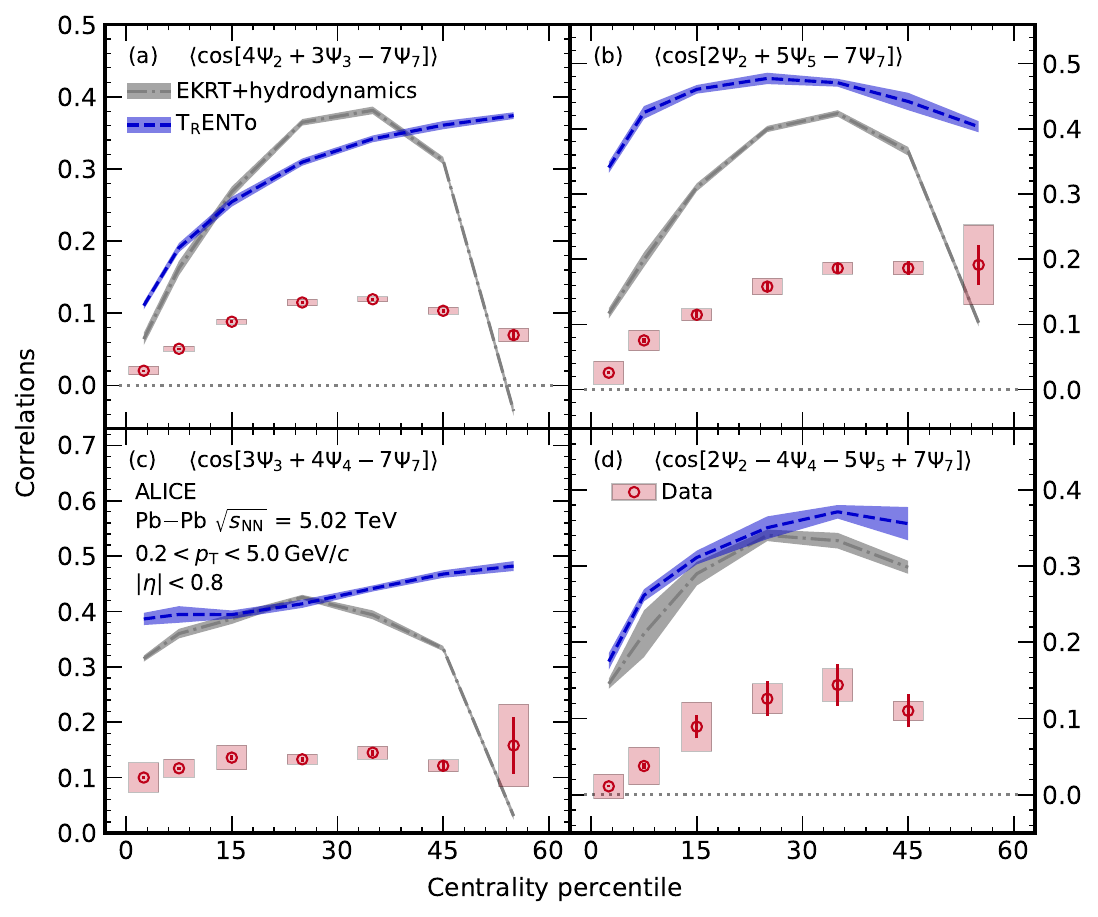}
    \caption{Comparison of the centrality dependence of the correlations between different combinations of SPC involving planes up to $\Psi_7$ (red circles) with the theoretical predictions from EKRT+hydrodynamics~\cite{Hirvonen:2022xfv} shown in gray bands. Initial-state predictions from \trento~\cite{Bass:1998ca, Bleicher:1999xi, Song:2007ux, Shen:2014vra} calculated with cumulant expansions are shown as blue bands. The lines (boxes) represent the statistical (systematic) uncertainties in the experimental data. The widths of the bands denote the statistical uncertainty of the model predictions.}
    \label{fig:hydro_fig6}
\end{figure}

\begin{figure}[t]
    \centering
    \includegraphics[width=\textwidth]{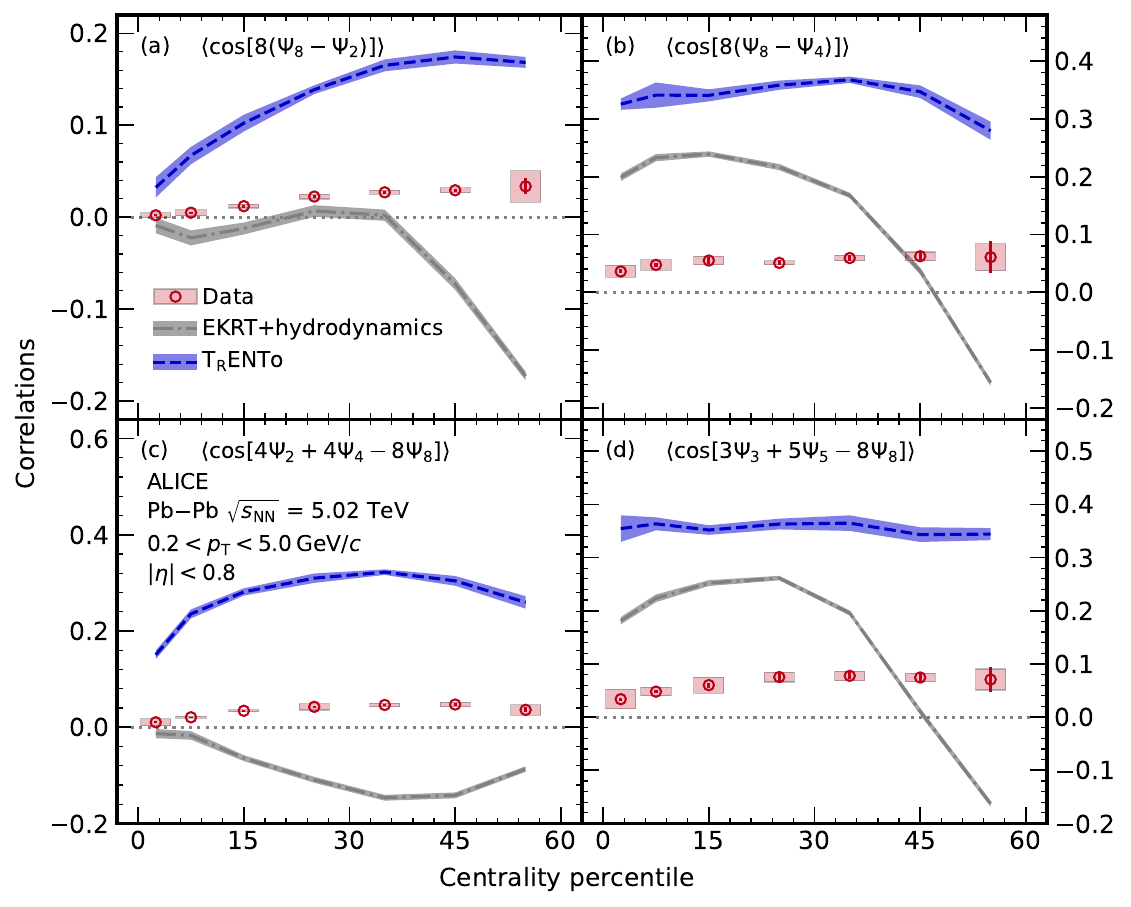}
    \caption{Comparison of the centrality dependence of the correlations between different combinations including $\Psi_8$ (red circles) with the theoretical predictions from EKRT+hydrodynamics~\cite{Hirvonen:2022xfv} shown in gray bands. Initial-state predictions from \trento~\cite{Bass:1998ca, Bleicher:1999xi, Song:2007ux, Shen:2014vra} calculated with cumulant expansions are shown as blue bands. The lines (boxes) represent the statistical (systematic) uncertainties of the experimental data. The widths of the bands denote the statistical uncertainty of the model predictions.}
    \label{fig:hydro_fig7}
\end{figure}

This Appendix presents observables not shown in Figs.~\ref{fig:hydro_fig1}--\ref{fig:hydro_fig4} compared with model calculations. Due to their large uncertainties, no final-state calculations from \trento+iEBE-VISHNU+UrQMD are presented in Figs.~\ref{fig:hydro_fig5}--\ref{fig:hydro_fig7}.

Additional combinations involving the symmetry planes $\Psi_2$, $\Psi_3$, $\Psi_4$, $\Psi_5$, and $\Psi_6$ can be seen in Fig.~\ref{fig:hydro_fig5}, including the first measured five-harmonic SPC among those planes.
Contrary to the previous observations, the predictions from EKRT+hydrodynamics not only overestimate the data but also do not manage to reproduce their centrality dependence. The complexity of the interplay between the non-linear response of $\Psi_5$ and $\Psi_6$ may be the cause of such a discrepancy. This confirms the need for higher-order observables to adjust the parameters in the theoretical models as well as the need for the improvement of the models themselves.
Furthermore, the comparison of the initial-state predictions from \trento~with the experimental data hints at a dampening of the initial state correlations, especially in more peripheral collisions. This might be due to a non-negligible non-linear response, which has to be addressed by future studies.

Figures~\ref{fig:hydro_fig6} and~\ref{fig:hydro_fig7} show results for four different SPC involving $\Psi_7$ and $\Psi_8$, respectively.
As for the results in Fig.~\ref{fig:hydro_fig5}, the predictions from EKRT+hydrodynamics~\cite{Hirvonen:2022xfv} cannot reproduce the centrality dependence of the experimental values. Interestingly, the calculations in Fig.~\ref{fig:hydro_fig6}~(a)--(c) and in Fig.~\ref{fig:hydro_fig7}~(a),~(b), and~(d) all present similar strongly decreasing signals for centrality percentiles above 40\% for the cases involving $\Psi_7$ and above 30\% of centrality for $\Psi_8$. In general, a decrease of correlation strength is expected in the transition from semicentral to peripheral collisions as the QCD medium produced becomes less thermalised. Thus, initial-state correlations are not properly transferred into the final-state momentum space, leading to the observed decrease in the correlation strength.
Note that the decease of correlation strength towards peripheral collisions seems in general to occur faster in the EKRT+hydrodynamics predictions than in the data, which could indicate a less thermalised medium in the model predictions.  As the predictions from EKRT+hydrodynamics are extracted at the hydrodynamic surface and, thus, do not contain separate hadronic transport models, further studies are required in that direction.

%============================================================%
\section{Effect of hadronic interactions}
\label{sec:app2}
\begin{figure}[t]
	\centering
	\includegraphics[width=\textwidth]{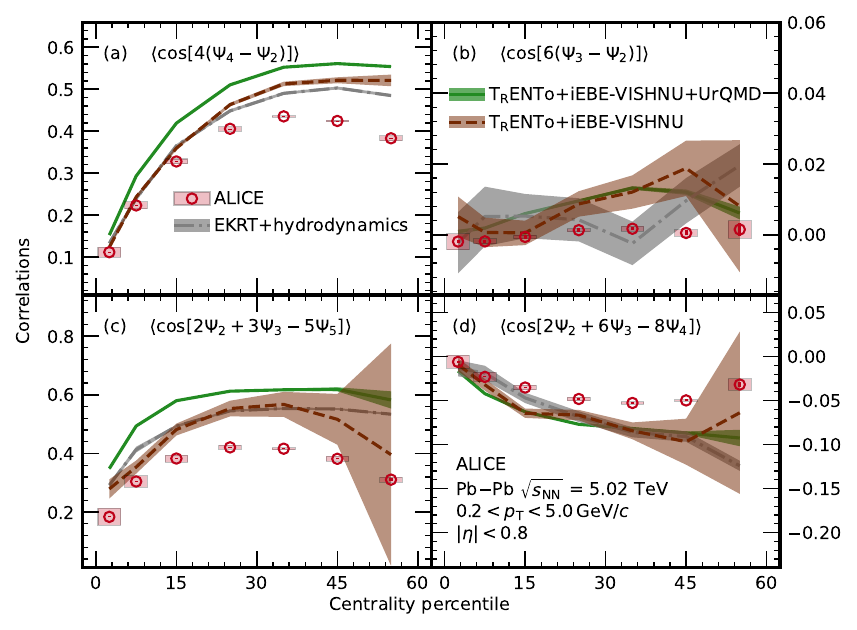}
	\caption{Comparison of predictions from models with and without hadronic cascade for a few selected SPC observables. The theoretical predictions from EKRT+hydrodynamics~\cite{Hirvonen:2022xfv} are shown in gray bands, \trento+iEBE-VISHNU+UrQMD in green bands, and \trento+iEBE-VISHNU in brown bands. The experimental data are shown as red circles with the lines (boxes) representing the statistical (systematic) uncertainties. The widths of the bands denote the statistical uncertainty of the model predictions.}
	\label{fig:nourqmd}
\end{figure}

In model comparison section (Sec.~\ref{sec:res}), a quantitative difference was found between EKRT+hydrodynamics and \trento+iEBE-VISHNU+UrQMD, which could be caused by a lack of hadronic afterburner model in the former. To quantify the effects of hadronic interactions on SPC, model set \trento+iEBE-VISHNU+UrQMD is run without UrQMD and the results are compared. Figure~\ref{fig:nourqmd} displays the model outcomes of \trento+iEBE-VISHNU+UrQMD and \trento+iEBE-VISHNU, which are compared with the EKRT+hydrodynamics and experimental results. A clear distinction in the magnitude between \trento+iEBE-VISHNU with and without UrQMD is shown for $\langle\cos[4(\Psi_4 - \Psi_2)]\rangle$ and $\langle\cos[2\Psi_2 + 3\Psi_3 -5\Psi_5]\rangle$. For these observables the \trento+iEBE-VISHNU predictions are of similar order with EKRT+hydrodynamics and, hence, in better agreement with the data. For $\langle\cos[6(\Psi_3 - \Psi_2)]\rangle$ and $\langle\cos[2\Psi_2 + 6\Psi_3 -8\Psi_4]\rangle$ the hadronic interactions do not increase the magnitude and the predictions from all three models are in a better agreement. As such, the difference between \trento+iEBE-VISHNU+UrQMD and EKRT+hydrodynamics can be partially explained with inclusion of a hadronic cascade model in the former. Nevertheless strong conclusions cannot be made because for \trento+iEBE-VISHNU+UrQMD and EKRT+hydrodynamics the model parameters are tuned to represent experimental data, mainly particle yields, anisotropic flow and mean transverse momentum distributions. Here \trento+iEBE-VISHNU predictions are evaluated with same parametrisation as for \trento+iEBE-VISHNU+UrQMD, and with that parametrisation the model does not reproduce the aforementioned experimental data. On average, the model produces fewer particles with higher transverse momentum when run without UrQMD. The parameter optimisation for the model without UrQMD is beyond the scope of this paper. Hadronic interactions play an important role in modelling heavy-ion collisions, but a more detailed analysis is needed to fully conclude their effects.
%

%============================================================%
\section{Comparison with results at lower beam energy}
\label{sec:app3}

\begin{figure}
    \centering
    \includegraphics[width=\textwidth]{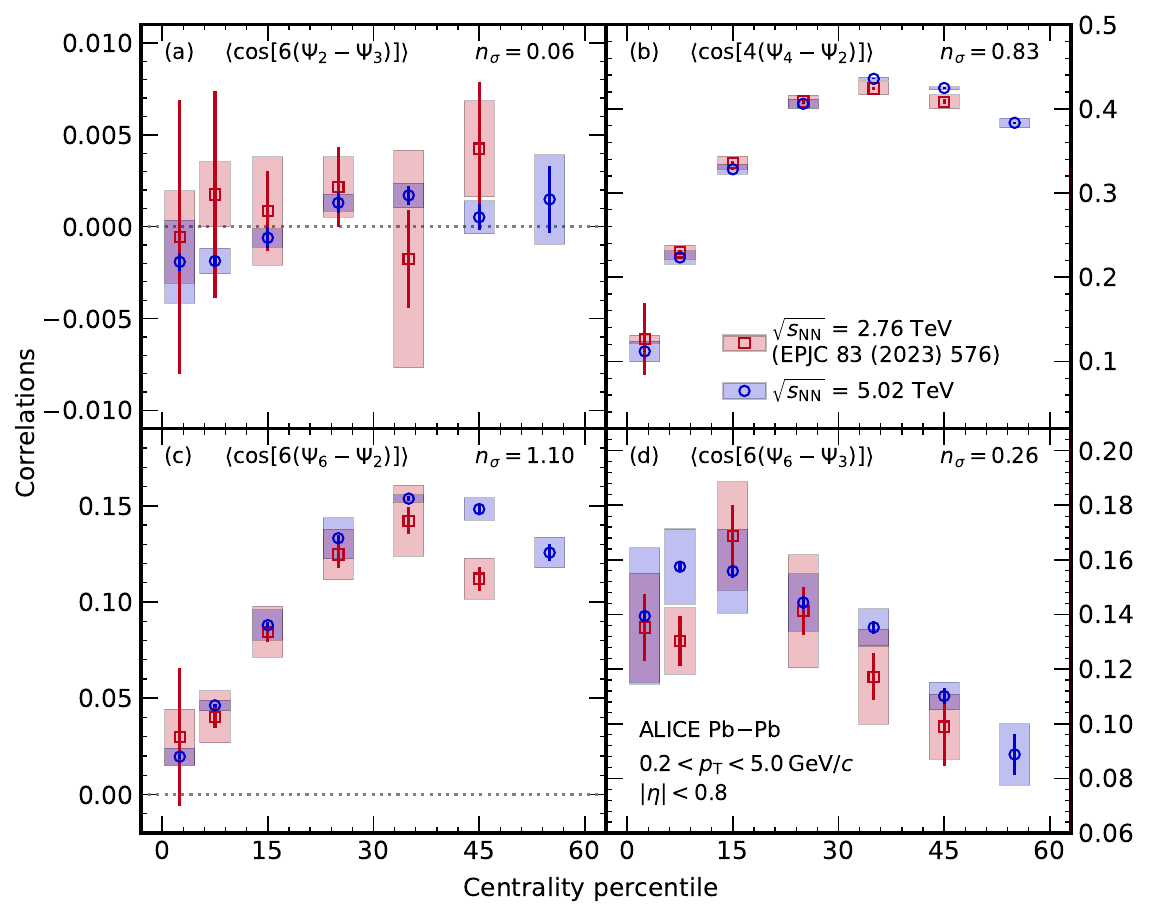}
    \caption{Correlations between two different symmetry planes measured at $\sqrt{s_{\rm NN}} = 2.76$~TeV (red squares) from~\cite{ALICE:2023wdn} and $\sqrt{s_{\rm NN}} = 5.02$~TeV (blue circles). The lines (boxes) represent the statistical (systematic) uncertainties. The agreement between the two sets of data is indicated with its number of $\sigma$ at the top right of each panel.}
    \label{fig:energy2SPC}
\end{figure}

\begin{figure}
    \centering
    \includegraphics[width=\textwidth]{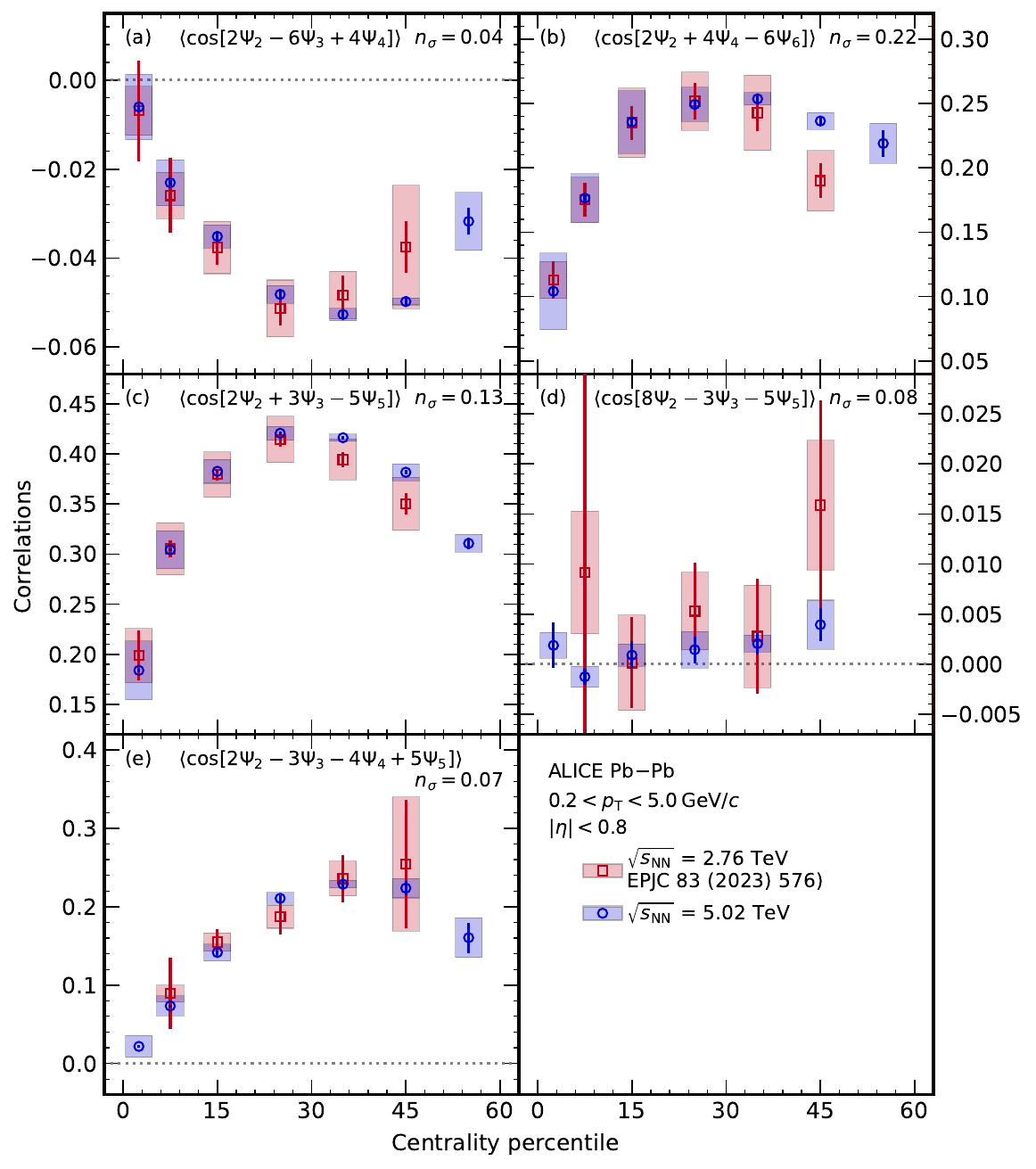}
    \caption{Correlations between three and four different symmetry planes measured at $\sqrt{s_{\rm NN}} = 2.76$~TeV (red squares) from~\cite{ALICE:2023wdn} and $\sqrt{s_{\rm NN}} = 5.02$~TeV (blue circles). The lines (boxes) represent the statistical (systematic) uncertainties. The agreement between the two sets of data is indicated with its number of $\sigma$ at the top right of each panel.}
    \label{fig:energy3SPC}
\end{figure}

Figures~\ref{fig:energy2SPC} and~\ref{fig:energy3SPC} display the comparison between the present results from Pb--Pb collisions at \fivenn and the SPC measured in Pb--Pb collisions at \twosevensixnn by ALICE~\cite{ALICE:2023wdn}. In Fig.~\ref{fig:energy2SPC}, the comparisons for the two-harmonic SPC $\langle \cos[6 (\Psi_2 - \Psi_3)] \rangle$, $\langle \cos[4 (\Psi_4 - \Psi_2)] \rangle$, $\langle \cos[6 (\Psi_6 - \Psi_2)] \rangle$, and $\langle \cos[6 (\Psi_6 - \Psi_3)] \rangle$ are shown.
With the current uncertainties, no significant deviation can be observed between the results at \twosevensixnn and \fivenn, and thus, the data do not support a strong energy dependence for those two-harmonic SPC. Compared with the results in Ref.~\cite{ALICE:2023wdn}, the present analysis manages to extend the upper centrality limit from 50\% to 60\%.
This extension of centrality range further indicates a decreasing correlation for $\langle \cos[4 (\Psi_4 - \Psi_2)] \rangle$, $\langle \cos[6 (\Psi_6 - \Psi_2)] \rangle$, and $\langle \cos[6 (\Psi_6 - \Psi_3)] \rangle$ in peripheral collisions. The reason for the observed decrease in the correlation strength is discussed in Appendix~\ref{sec:app1}. The SPC $\langle \cos[6 (\Psi_2 - \Psi_3)] \rangle$ was found to be compatible with zero within the uncertainties for all centrality intervals in the analysis at \twosevensixnn by ALICE~\cite{ALICE:2023wdn}. With the increased precision of the present analysis, the correlation signal of $\langle \cos[6 (\Psi_2 - \Psi_3)] \rangle$ is still compatible with zero within $n_\sigma = 2.1$. The interpretation of this absence of correlation is discussed in Sec.~\ref{subsec:hydro}.

Similarly, Fig.~\ref{fig:energy3SPC} presents the energy dependence for three- and four-harmonic SPC. As in the case of the two-harmonic SPC, no significant energy dependence can be observed within the current uncertainties. Except for the three-harmonic SPC $\langle \cos[8\Psi_2 - 3\Psi_3 - 5\Psi_5] \rangle$,  the present analysis manages to extend also the three- and four-harmonic measurements of Ref.~\cite{ALICE:2023wdn} up to 60\% centrality. For the two SPC $\langle \cos[8\Psi_2 - 3\Psi_3 - 5\Psi_5] \rangle$ and $\langle \cos[2\Psi_2 - 3\Psi_3 - 4\Psi_4 + 5\Psi_5] \rangle$, correlation signals are extracted in the 0--5\% centrality for the first time as well. 
Overall, this extension adds to the available information of SPC in Pb--Pb collisions at LHC energies. Furthermore, the results at \fivenn reinforce the observations from Ref.~\cite{ALICE:2023wdn} of $\langle \cos[8\Psi_2 - 3\Psi_3 - 5\Psi_5] \rangle$ being compatible with zero in the whole centrality range ($n_\sigma = 0.82$ significance).

\clearpage

%%%%% Authorlist - please do not touch: handled by EB chairs 
\section{The ALICE Collaboration}
\label{app:collab}
\input{Alice_Authorlist_2024-08-10_Opt_C.tex}

\end{document}

%% file: commands.tex
%%%%%%%%%%%%%%%%%%%%%%%%%%%%%%%%%%%%%%%%%%%%%%%%%%
% These are some new commands that may be useful 
% for paper writing in general. If other newcommands
% are needed for your specific paper, please feel 
% free to add here. 
%
% The currently available commands are organized in: 
% 1) Systems
% 2) Quantities
% 3) Energies and units
% 4) Detectors
% 5) particle species 
%%%%%%%%%%%%%%%%%%%%%%%%%%%%%%%%%%%%%%%%%%%%%%%%%%

% 1) SYSTEMS 
\newcommand{\pp}           {pp\xspace}
\newcommand{\ppbar}        {\mbox{$\mathrm {p\overline{p}}$}\xspace}
\newcommand{\XeXe}         {\mbox{Xe--Xe}\xspace}
\newcommand{\PbPb}         {\mbox{Pb--Pb}\xspace}
\newcommand{\pA}           {\mbox{pA}\xspace}
\newcommand{\pPb}          {\mbox{p--Pb}\xspace}
\newcommand{\AuAu}         {\mbox{Au--Au}\xspace}
\newcommand{\dAu}          {\mbox{d--Au}\xspace}

% 2) QUANTITIES 
\newcommand{\s}            {\ensuremath{\sqrt{s}}\xspace}
\newcommand{\snn}          {\ensuremath{\sqrt{s_{\mathrm{NN}}}}\xspace}
\newcommand{\pt}           {\ensuremath{p_{\rm T}}\xspace}
\newcommand{\meanpt}       {$\langle p_{\mathrm{T}}\rangle$\xspace}
\newcommand{\ycms}         {\ensuremath{y_{\rm CMS}}\xspace}
\newcommand{\ylab}         {\ensuremath{y_{\rm lab}}\xspace}
\newcommand{\etarange}[1]  {\mbox{$\left | \eta \right |~<~#1$}}
\newcommand{\yrange}[1]    {\mbox{$\left | y \right |~<~#1$}}
\newcommand{\dndy}         {\ensuremath{\mathrm{d}N_\mathrm{ch}/\mathrm{d}y}\xspace}
\newcommand{\dndeta}       {\ensuremath{\mathrm{d}N_\mathrm{ch}/\mathrm{d}\eta}\xspace}
\newcommand{\avdndeta}     {\ensuremath{\langle\dndeta\rangle}\xspace}
\newcommand{\dNdy}         {\ensuremath{\mathrm{d}N_\mathrm{ch}/\mathrm{d}y}\xspace}
\newcommand{\Npart}        {\ensuremath{N_\mathrm{part}}\xspace}
\newcommand{\Ncoll}        {\ensuremath{N_\mathrm{coll}}\xspace}
\newcommand{\dEdx}         {\ensuremath{\textrm{d}E/\textrm{d}x}\xspace}
\newcommand{\RpPb}         {\ensuremath{R_{\rm pPb}}\xspace}

% 3) ENERGIES, UNITS
\newcommand{\nineH}        {$\sqrt{s}~=~0.9$~Te\kern-.1emV\xspace}
\newcommand{\seven}        {$\sqrt{s}~=~7$~Te\kern-.1emV\xspace}
\newcommand{\twoH}         {$\sqrt{s}~=~0.2$~Te\kern-.1emV\xspace}
\newcommand{\twosevensix}  {$\sqrt{s}~=~2.76$~Te\kern-.1emV\xspace}
\newcommand{\five}         {$\sqrt{s}~=~5.02$~Te\kern-.1emV\xspace}
\newcommand{\twosevensixnn}{$\sqrt{s_{\mathrm{NN}}}~=~2.76$~Te\kern-.1emV\xspace}
\newcommand{\fivenn}       {$\sqrt{s_{\mathrm{NN}}}~=~5.02$~Te\kern-.1emV\xspace}
\newcommand{\LT}           {L{\'e}vy-Tsallis\xspace}
\newcommand{\GeVc}         {Ge\kern-.1emV/$c$\xspace}
\newcommand{\MeVc}         {Me\kern-.1emV/$c$\xspace}
\newcommand{\TeV}          {Te\kern-.1emV\xspace}
\newcommand{\GeV}          {Ge\kern-.1emV\xspace}
\newcommand{\MeV}          {Me\kern-.1emV\xspace}
\newcommand{\GeVmass}      {Ge\kern-.2emV/$c^2$\xspace}
\newcommand{\MeVmass}      {Me\kern-.2emV/$c^2$\xspace}
\newcommand{\lumi}         {\ensuremath{\mathcal{L}}\xspace}

% 4) DETECTORS 
\newcommand{\ITS}          {\rm{ITS}\xspace}
\newcommand{\TOF}          {\rm{TOF}\xspace}
\newcommand{\ZDC}          {\rm{ZDC}\xspace}
\newcommand{\ZDCs}         {\rm{ZDCs}\xspace}
\newcommand{\ZNA}          {\rm{ZNA}\xspace}
\newcommand{\ZNC}          {\rm{ZNC}\xspace}
\newcommand{\SPD}          {\rm{SPD}\xspace}
\newcommand{\SDD}          {\rm{SDD}\xspace}
\newcommand{\SSD}          {\rm{SSD}\xspace}
\newcommand{\TPC}          {\rm{TPC}\xspace}
\newcommand{\TRD}          {\rm{TRD}\xspace}
\newcommand{\VZERO}        {\rm{V0}\xspace}
\newcommand{\VZEROA}       {\rm{V0A}\xspace}
\newcommand{\VZEROC}       {\rm{V0C}\xspace}
\newcommand{\Vdecay} 	   {\ensuremath{V^{0}}\xspace}

% 4) PARTICLE SPECIES 
\newcommand{\ee}           {\ensuremath{e^{+}e^{-}}} 
\newcommand{\pip}          {\ensuremath{\pi^{+}}\xspace}
\newcommand{\pim}          {\ensuremath{\pi^{-}}\xspace}
\newcommand{\kap}          {\ensuremath{\rm{K}^{+}}\xspace}
\newcommand{\kam}          {\ensuremath{\rm{K}^{-}}\xspace}
\newcommand{\pbar}         {\ensuremath{\rm\overline{p}}\xspace}
\newcommand{\kzero}        {\ensuremath{{\rm K}^{0}_{\rm{S}}}\xspace}
\newcommand{\lmb}          {\ensuremath{\Lambda}\xspace}
\newcommand{\almb}         {\ensuremath{\overline{\Lambda}}\xspace}
\newcommand{\Om}           {\ensuremath{\Omega^-}\xspace}
\newcommand{\Mo}           {\ensuremath{\overline{\Omega}^+}\xspace}
\newcommand{\X}            {\ensuremath{\Xi^-}\xspace}
\newcommand{\Ix}           {\ensuremath{\overline{\Xi}^+}\xspace}
\newcommand{\Xis}          {\ensuremath{\Xi^{\pm}}\xspace}
\newcommand{\Oms}          {\ensuremath{\Omega^{\pm}}\xspace}
\newcommand{\degree}       {\ensuremath{^{\rm o}}\xspace}

% 5) Additional
\newcommand{\trento}       {T$_{\mathrm R}$ENTo}

%% file: fa_2024-08-10_Opt_C.tex
% Version: 2024-08-10

The ALICE Collaboration would like to thank all its engineers and technicians for their invaluable contributions to the construction of the experiment and the CERN accelerator teams for the outstanding performance of the LHC complex.
The ALICE Collaboration gratefully acknowledges the resources and support provided by all Grid centres and the Worldwide LHC Computing Grid (WLCG) collaboration.
The ALICE Collaboration acknowledges the following funding agencies for their support in building and running the ALICE detector:
A. I. Alikhanyan National Science Laboratory (Yerevan Physics Institute) Foundation (ANSL), State Committee of Science and World Federation of Scientists (WFS), Armenia;
Austrian Academy of Sciences, Austrian Science Fund (FWF): [M 2467-N36] and Nationalstiftung f\"{u}r Forschung, Technologie und Entwicklung, Austria;
Ministry of Communications and High Technologies, National Nuclear Research Center, Azerbaijan;
Conselho Nacional de Desenvolvimento Cient\'{\i}fico e Tecnol\'{o}gico (CNPq), Financiadora de Estudos e Projetos (Finep), Funda\c{c}\~{a}o de Amparo \`{a} Pesquisa do Estado de S\~{a}o Paulo (FAPESP) and Universidade Federal do Rio Grande do Sul (UFRGS), Brazil;
Bulgarian Ministry of Education and Science, within the National Roadmap for Research Infrastructures 2020-2027 (object CERN), Bulgaria;
Ministry of Education of China (MOEC) , Ministry of Science \& Technology of China (MSTC) and National Natural Science Foundation of China (NSFC), China;
Ministry of Science and Education and Croatian Science Foundation, Croatia;
Centro de Aplicaciones Tecnol\'{o}gicas y Desarrollo Nuclear (CEADEN), Cubaenerg\'{\i}a, Cuba;
Ministry of Education, Youth and Sports of the Czech Republic, Czech Republic;
The Danish Council for Independent Research | Natural Sciences, the VILLUM FONDEN and Danish National Research Foundation (DNRF), Denmark;
Helsinki Institute of Physics (HIP), Finland;
Commissariat \`{a} l'Energie Atomique (CEA) and Institut National de Physique Nucl\'{e}aire et de Physique des Particules (IN2P3) and Centre National de la Recherche Scientifique (CNRS), France;
Bundesministerium f\"{u}r Bildung und Forschung (BMBF) and GSI Helmholtzzentrum f\"{u}r Schwerionenforschung GmbH, Germany;
General Secretariat for Research and Technology, Ministry of Education, Research and Religions, Greece;
National Research, Development and Innovation Office, Hungary;
Department of Atomic Energy Government of India (DAE), Department of Science and Technology, Government of India (DST), University Grants Commission, Government of India (UGC) and Council of Scientific and Industrial Research (CSIR), India;
National Research and Innovation Agency - BRIN, Indonesia;
Istituto Nazionale di Fisica Nucleare (INFN), Italy;
Japanese Ministry of Education, Culture, Sports, Science and Technology (MEXT) and Japan Society for the Promotion of Science (JSPS) KAKENHI, Japan;
Consejo Nacional de Ciencia (CONACYT) y Tecnolog\'{i}a, through Fondo de Cooperaci\'{o}n Internacional en Ciencia y Tecnolog\'{i}a (FONCICYT) and Direcci\'{o}n General de Asuntos del Personal Academico (DGAPA), Mexico;
Nederlandse Organisatie voor Wetenschappelijk Onderzoek (NWO), Netherlands;
The Research Council of Norway, Norway;
Pontificia Universidad Cat\'{o}lica del Per\'{u}, Peru;
Ministry of Science and Higher Education, National Science Centre and WUT ID-UB, Poland;
Korea Institute of Science and Technology Information and National Research Foundation of Korea (NRF), Republic of Korea;
Ministry of Education and Scientific Research, Institute of Atomic Physics, Ministry of Research and Innovation and Institute of Atomic Physics and Universitatea Nationala de Stiinta si Tehnologie Politehnica Bucuresti, Romania;
Ministry of Education, Science, Research and Sport of the Slovak Republic, Slovakia;
National Research Foundation of South Africa, South Africa;
Swedish Research Council (VR) and Knut \& Alice Wallenberg Foundation (KAW), Sweden;
European Organization for Nuclear Research, Switzerland;
Suranaree University of Technology (SUT), National Science and Technology Development Agency (NSTDA) and National Science, Research and Innovation Fund (NSRF via PMU-B B05F650021), Thailand;
Turkish Energy, Nuclear and Mineral Research Agency (TENMAK), Turkey;
National Academy of  Sciences of Ukraine, Ukraine;
Science and Technology Facilities Council (STFC), United Kingdom;
National Science Foundation of the United States of America (NSF) and United States Department of Energy, Office of Nuclear Physics (DOE NP), United States of America.
In addition, individual groups or members have received support from:
Czech Science Foundation (grant no. 23-07499S), Czech Republic;
FORTE project, reg.\ no.\ CZ.02.01.01/00/22\_008/0004632, Czech Republic, co-funded by the European Union, Czech Republic;
European Research Council (grant no. 950692), European Union;
ICSC - Centro Nazionale di Ricerca in High Performance Computing, Big Data and Quantum Computing, European Union - NextGenerationEU;
Academy of Finland (Center of Excellence in Quark Matter) (grant nos. 346327, 346328), Finland.

%% file: Alice_Authorlist_2024-08-10_Opt_C.tex
% ALICE Collaboration author list for 2024-08-10
\begin{flushleft} 
\small

S.~Acharya\,\orcidlink{0000-0002-9213-5329}\,$^{\rm 127}$, 
A.~Agarwal$^{\rm 135}$, 
G.~Aglieri Rinella\,\orcidlink{0000-0002-9611-3696}\,$^{\rm 32}$, 
L.~Aglietta\,\orcidlink{0009-0003-0763-6802}\,$^{\rm 24}$, 
M.~Agnello\,\orcidlink{0000-0002-0760-5075}\,$^{\rm 29}$, 
N.~Agrawal\,\orcidlink{0000-0003-0348-9836}\,$^{\rm 25}$, 
Z.~Ahammed\,\orcidlink{0000-0001-5241-7412}\,$^{\rm 135}$, 
S.~Ahmad\,\orcidlink{0000-0003-0497-5705}\,$^{\rm 15}$, 
S.U.~Ahn\,\orcidlink{0000-0001-8847-489X}\,$^{\rm 71}$, 
I.~Ahuja\,\orcidlink{0000-0002-4417-1392}\,$^{\rm 37}$
% ...

\section*{Affiliation Notes}

$^{\rm I}$ Deceased\\
$^{\rm II}$ Also at: Max-Planck-Institut fur Physik, Munich, Germany\\
$^{\rm III}$ Also at: Italian National Agency for New Technologies, Energy and Sustainable Economic Development (ENEA), Bologna, Italy\\
$^{\rm IV}$ Also at: Dipartimento DET del Politecnico di Torino, Turin, Italy\\
$^{\rm V}$ Also at: Yildiz Technical University, Istanbul, T\"{u}rkiye\\
$^{\rm VI}$ Also at: Department of Applied Physics, Aligarh Muslim University, Aligarh, India\\
$^{\rm VII}$ Also at: Institute of Theoretical Physics, University of Wroclaw, Poland\\
$^{\rm VIII}$ Also at: An institution covered by a cooperation agreement with CERN\\

\section*{Collaboration Institutes}

$^{1}$ A.I. Alikhanyan National Science Laboratory (Yerevan Physics Institute) Foundation, Yerevan, Armenia\\
$^{2}$ AGH University of Krakow, Cracow, Poland\\
$^{3}$ Bogolyubov Institute for Theoretical Physics, National Academy of Sciences of Ukraine, Kiev, Ukraine\\
$^{4}$ Bose Institute, Department of Physics  and Centre for Astroparticle Physics and Space Science (CAPSS), Kolkata, India\\
$^{5}$ California Polytechnic State University, San Luis Obispo, California, United States\\
$^{6}$ Central China Normal University, Wuhan, China\\
$^{7}$ Centro de Aplicaciones Tecnol\'{o}gicas y Desarrollo Nuclear (CEADEN), Havana, Cuba\\
$^{8}$ Centro de Investigaci\'{o}n y de Estudios Avanzados (CINVESTAV), Mexico City and M\'{e}rida, Mexico\\
$^{9}$ Chicago State University, Chicago, Illinois, United States\\
$^{10}$ China Institute of Atomic Energy, Beijing, China\\
$^{11}$ China University of Geosciences, Wuhan, China\\
$^{12}$ Chungbuk National University, Cheongju, Republic of Korea\\
$^{13}$ Comenius University Bratislava, Faculty of Mathematics, Physics and Informatics, Bratislava, Slovak Republic\\
$^{14}$ Creighton University, Omaha, Nebraska, United States\\
$^{15}$ Department of Physics, Aligarh Muslim University, Aligarh, India\\
$^{16}$ Department of Physics, Pusan National University, Pusan, Republic of Korea\\
$^{17}$ Department of Physics, Sejong University, Seoul, Republic of Korea\\
$^{18}$ Department of Physics, University of California, Berkeley, California, United States\\
$^{19}$ Department of Physics, University of Oslo, Oslo, Norway\\
$^{20}$ Department of Physics and Technology, University of Bergen, Bergen, Norway\\
$^{21}$ Dipartimento di Fisica, Universit\`{a} di Pavia, Pavia, Italy\\
$^{22}$ Dipartimento di Fisica dell'Universit\`{a} and Sezione INFN, Cagliari, Italy\\
$^{23}$ Dipartimento di Fisica dell'Universit\`{a} and Sezione INFN, Trieste, Italy\\
$^{24}$ Dipartimento di Fisica dell'Universit\`{a} and Sezione INFN, Turin, Italy\\
$^{25}$ Dipartimento di Fisica e Astronomia dell'Universit\`{a} and Sezione INFN, Bologna, Italy\\
$^{26}$ Dipartimento di Fisica e Astronomia dell'Universit\`{a} and Sezione INFN, Catania, Italy\\
$^{27}$ Dipartimento di Fisica e Astronomia dell'Universit\`{a} and Sezione INFN, Padova, Italy\\
$^{28}$ Dipartimento di Fisica `E.R.~Caianiello' dell'Universit\`{a} and Gruppo Collegato INFN, Salerno, Italy\\
$^{29}$ Dipartimento DISAT del Politecnico and Sezione INFN, Turin, Italy\\
$^{30}$ Dipartimento di Scienze MIFT, Universit\`{a} di Messina, Messina, Italy\\
$^{31}$ Dipartimento Interateneo di Fisica `M.~Merlin' and Sezione INFN, Bari, Italy\\
$^{32}$ European Organization for Nuclear Research (CERN), Geneva, Switzerland\\
$^{33}$ Faculty of Electrical Engineering, Mechanical Engineering and Naval Architecture, University of Split, Split, Croatia\\
$^{34}$ Faculty of Engineering and Science, Western Norway University of Applied Sciences, Bergen, Norway\\
$^{35}$ Faculty of Nuclear Sciences and Physical Engineering, Czech Technical University in Prague, Prague, Czech Republic\\
$^{36}$ Faculty of Physics, Sofia University, Sofia, Bulgaria\\
$^{37}$ Faculty of Science, P.J.~\v{S}af\'{a}rik University, Ko\v{s}ice, Slovak Republic\\
$^{38}$ Frankfurt Institute for Advanced Studies, Johann Wolfgang Goethe-Universit\"{a}t Frankfurt, Frankfurt, Germany\\
$^{39}$ Fudan University, Shanghai, China\\
$^{40}$ Gangneung-Wonju National University, Gangneung, Republic of Korea\\
$^{41}$ Gauhati University, Department of Physics, Guwahati, India\\
$^{42}$ Helmholtz-Institut f\"{u}r Strahlen- und Kernphysik, Rheinische Friedrich-Wilhelms-Universit\"{a}t Bonn, Bonn, Germany\\
$^{43}$ Helsinki Institute of Physics (HIP), Helsinki, Finland\\
$^{44}$ High Energy Physics Group,  Universidad Aut\'{o}noma de Puebla, Puebla, Mexico\\
$^{45}$ Horia Hulubei National Institute of Physics and Nuclear Engineering, Bucharest, Romania\\
$^{46}$ HUN-REN Wigner Research Centre for Physics, Budapest, Hungary\\
$^{47}$ Indian Institute of Technology Bombay (IIT), Mumbai, India\\
$^{48}$ Indian Institute of Technology Indore, Indore, India\\
$^{49}$ INFN, Laboratori Nazionali di Frascati, Frascati, Italy\\
$^{50}$ INFN, Sezione di Bari, Bari, Italy\\
$^{51}$ INFN, Sezione di Bologna, Bologna, Italy\\
$^{52}$ INFN, Sezione di Cagliari, Cagliari, Italy\\
$^{53}$ INFN, Sezione di Catania, Catania, Italy\\
$^{54}$ INFN, Sezione di Padova, Padova, Italy\\
$^{55}$ INFN, Sezione di Pavia, Pavia, Italy\\
$^{56}$ INFN, Sezione di Torino, Turin, Italy\\
$^{57}$ INFN, Sezione di Trieste, Trieste, Italy\\
$^{58}$ Inha University, Incheon, Republic of Korea\\
$^{59}$ Institute for Gravitational and Subatomic Physics (GRASP), Utrecht University/Nikhef, Utrecht, Netherlands\\
$^{60}$ Institute of Experimental Physics, Slovak Academy of Sciences, Ko\v{s}ice, Slovak Republic\\
$^{61}$ Institute of Physics, Homi Bhabha National Institute, Bhubaneswar, India\\
$^{62}$ Institute of Physics of the Czech Academy of Sciences, Prague, Czech Republic\\
$^{63}$ Institute of Space Science (ISS), Bucharest, Romania\\
$^{64}$ Institut f\"{u}r Kernphysik, Johann Wolfgang Goethe-Universit\"{a}t Frankfurt, Frankfurt, Germany\\
$^{65}$ Instituto de Ciencias Nucleares, Universidad Nacional Aut\'{o}noma de M\'{e}xico, Mexico City, Mexico\\
$^{66}$ Instituto de F\'{i}sica, Universidade Federal do Rio Grande do Sul (UFRGS), Porto Alegre, Brazil\\
$^{67}$ Instituto de F\'{\i}sica, Universidad Nacional Aut\'{o}noma de M\'{e}xico, Mexico City, Mexico\\
$^{68}$ iThemba LABS, National Research Foundation, Somerset West, South Africa\\
$^{69}$ Jeonbuk National University, Jeonju, Republic of Korea\\
$^{70}$ Johann-Wolfgang-Goethe Universit\"{a}t Frankfurt Institut f\"{u}r Informatik, Fachbereich Informatik und Mathematik, Frankfurt, Germany\\
$^{71}$ Korea Institute of Science and Technology Information, Daejeon, Republic of Korea\\
$^{72}$ KTO Karatay University, Konya, Turkey\\
$^{73}$ Laboratoire de Physique Subatomique et de Cosmologie, Universit\'{e} Grenoble-Alpes, CNRS-IN2P3, Grenoble, France\\
$^{74}$ Lawrence Berkeley National Laboratory, Berkeley, California, United States\\
$^{75}$ Lund University Department of Physics, Division of Particle Physics, Lund, Sweden\\
$^{76}$ Nagasaki Institute of Applied Science, Nagasaki, Japan\\
$^{77}$ Nara Women{'}s University (NWU), Nara, Japan\\
$^{78}$ National and Kapodistrian University of Athens, School of Science, Department of Physics , Athens, Greece\\
$^{79}$ National Centre for Nuclear Research, Warsaw, Poland\\
$^{80}$ National Institute of Science Education and Research, Homi Bhabha National Institute, Jatni, India\\
$^{81}$ National Nuclear Research Center, Baku, Azerbaijan\\
$^{82}$ National Research and Innovation Agency - BRIN, Jakarta, Indonesia\\
$^{83}$ Niels Bohr Institute, University of Copenhagen, Copenhagen, Denmark\\
$^{84}$ Nikhef, National institute for subatomic physics, Amsterdam, Netherlands\\
$^{85}$ Nuclear Physics Group, STFC Daresbury Laboratory, Daresbury, United Kingdom\\
$^{86}$ Nuclear Physics Institute of the Czech Academy of Sciences, Husinec-\v{R}e\v{z}, Czech Republic\\
$^{87}$ Oak Ridge National Laboratory, Oak Ridge, Tennessee, United States\\
$^{88}$ Ohio State University, Columbus, Ohio, United States\\
$^{89}$ Physics department, Faculty of science, University of Zagreb, Zagreb, Croatia\\
$^{90}$ Physics Department, Panjab University, Chandigarh, India\\
$^{91}$ Physics Department, University of Jammu, Jammu, India\\
$^{92}$ Physics Program and International Institute for Sustainability with Knotted Chiral Meta Matter (SKCM2), Hiroshima University, Hiroshima, Japan\\
$^{93}$ Physikalisches Institut, Eberhard-Karls-Universit\"{a}t T\"{u}bingen, T\"{u}bingen, Germany\\
$^{94}$ Physikalisches Institut, Ruprecht-Karls-Universit\"{a}t Heidelberg, Heidelberg, Germany\\
$^{95}$ Physik Department, Technische Universit\"{a}t M\"{u}nchen, Munich, Germany\\
$^{96}$ Politecnico di Bari and Sezione INFN, Bari, Italy\\
$^{97}$ Research Division and ExtreMe Matter Institute EMMI, GSI Helmholtzzentrum f\"ur Schwerionenforschung GmbH, Darmstadt, Germany\\
$^{98}$ Saga University, Saga, Japan\\
$^{99}$ Saha Institute of Nuclear Physics, Homi Bhabha National Institute, Kolkata, India\\
$^{100}$ School of Physics and Astronomy, University of Birmingham, Birmingham, United Kingdom\\
$^{101}$ Secci\'{o}n F\'{\i}sica, Departamento de Ciencias, Pontificia Universidad Cat\'{o}lica del Per\'{u}, Lima, Peru\\
$^{102}$ Stefan Meyer Institut f\"{u}r Subatomare Physik (SMI), Vienna, Austria\\
$^{103}$ SUBATECH, IMT Atlantique, Nantes Universit\'{e}, CNRS-IN2P3, Nantes, France\\
$^{104}$ Sungkyunkwan University, Suwon City, Republic of Korea\\
$^{105}$ Suranaree University of Technology, Nakhon Ratchasima, Thailand\\
$^{106}$ Technical University of Ko\v{s}ice, Ko\v{s}ice, Slovak Republic\\
$^{107}$ The Henryk Niewodniczanski Institute of Nuclear Physics, Polish Academy of Sciences, Cracow, Poland\\
$^{108}$ The University of Texas at Austin, Austin, Texas, United States\\
$^{109}$ Universidad Aut\'{o}noma de Sinaloa, Culiac\'{a}n, Mexico\\
$^{110}$ Universidade de S\~{a}o Paulo (USP), S\~{a}o Paulo, Brazil\\
$^{111}$ Universidade Estadual de Campinas (UNICAMP), Campinas, Brazil\\
$^{112}$ Universidade Federal do ABC, Santo Andre, Brazil\\
$^{113}$ Universitatea Nationala de Stiinta si Tehnologie Politehnica Bucuresti, Bucharest, Romania\\
$^{114}$ University of Cape Town, Cape Town, South Africa\\
$^{115}$ University of Derby, Derby, United Kingdom\\
$^{116}$ University of Houston, Houston, Texas, United States\\
$^{117}$ University of Jyv\"{a}skyl\"{a}, Jyv\"{a}skyl\"{a}, Finland\\
$^{118}$ University of Kansas, Lawrence, Kansas, United States\\
$^{119}$ University of Liverpool, Liverpool, United Kingdom\\
$^{120}$ University of Science and Technology of China, Hefei, China\\
$^{121}$ University of South-Eastern Norway, Kongsberg, Norway\\
$^{122}$ University of Tennessee, Knoxville, Tennessee, United States\\
$^{123}$ University of the Witwatersrand, Johannesburg, South Africa\\
$^{124}$ University of Tokyo, Tokyo, Japan\\
$^{125}$ University of Tsukuba, Tsukuba, Japan\\
$^{126}$ Universit\"{a}t M\"{u}nster, Institut f\"{u}r Kernphysik, M\"{u}nster, Germany\\
$^{127}$ Universit\'{e} Clermont Auvergne, CNRS/IN2P3, LPC, Clermont-Ferrand, France\\
$^{128}$ Universit\'{e} de Lyon, CNRS/IN2P3, Institut de Physique des 2 Infinis de Lyon, Lyon, France\\
$^{129}$ Universit\'{e} de Strasbourg, CNRS, IPHC UMR 7178, F-67000 Strasbourg, France, Strasbourg, France\\
$^{130}$ Universit\'{e} Paris-Saclay, Centre d'Etudes de Saclay (CEA), IRFU, D\'{e}partment de Physique Nucl\'{e}aire (DPhN), Saclay, France\\
$^{131}$ Universit\'{e}  Paris-Saclay, CNRS/IN2P3, IJCLab, Orsay, France\\
$^{132}$ Universit\`{a} degli Studi di Foggia, Foggia, Italy\\
$^{133}$ Universit\`{a} del Piemonte Orientale, Vercelli, Italy\\
$^{134}$ Universit\`{a} di Brescia, Brescia, Italy\\
$^{135}$ Variable Energy Cyclotron Centre, Homi Bhabha National Institute, Kolkata, India\\
$^{136}$ Warsaw University of Technology, Warsaw, Poland\\
$^{137}$ Wayne State University, Detroit, Michigan, United States\\
$^{138}$ Yale University, New Haven, Connecticut, United States\\
$^{139}$ Yonsei University, Seoul, Republic of Korea\\
$^{140}$  Zentrum  f\"{u}r Technologie und Transfer (ZTT), Worms, Germany\\
$^{141}$ Affiliated with an institute covered by a cooperation agreement with CERN\\
$^{142}$ Affiliated with an international laboratory covered by a cooperation agreement with CERN.\\

\end{flushleft}